\documentclass[11 pt, onecolumn]{article}
\usepackage[margin=1.1in]{geometry}

\usepackage{amsmath}
\usepackage{bm}	
\usepackage{graphicx}

\usepackage{graphicx}
\usepackage{multirow}
\usepackage{mdwlist}

\usepackage{threeparttable}

\usepackage{amsfonts}
\usepackage{amsmath}
\usepackage{hyperref}
\usepackage{bbm}
\usepackage{float}
\usepackage{caption}
\usepackage{subcaption}
\usepackage{xcolor}
\usepackage{mathtools}
\usepackage{tikz}

\usetikzlibrary{shapes.geometric,arrows}
\usepackage{enumerate}
\usepackage{multicol}
\usepackage{stackrel}

\usepackage{subcaption}
\usepackage{amssymb}
\usepackage[utf8]{inputenc}
\usepackage{booktabs,lipsum}
\usepackage{hyperref}
\usepackage{algorithm}
\usepackage{algpseudocode}
\usepackage{caption}

\usepackage{longtable}

\DeclareMathAlphabet{\mathcal}{OMS}{cmsy}{m}{n}

\usepackage{graphicx}
\usepackage{epstopdf}

\begin{document}
\title{\LARGE \bf Transforming Medical Regulations into Numbers: Vectorizing a Decade of Medical Device Regulatory Shifts in the USA, EU, and China}

\author{Yu Han, Jeroen H.M. Bergmann}

\newcommand*{\QEDA}{\hfill\ensuremath{\blacksquare}}%

\maketitle

\begin{abstract}
Navigating the regulatory frameworks that ensure the safety and efficacy of medical devices can be challenging, especially across different regions. These frameworks often require redundant testing, slowing down the process of getting innovations to patients. This study leverages Natural Language Processing (NLP) to analyze 664 regulations and guidelines from the USA, EU, and China over the past decade, covering over 200 million tokens. We categorize regulations into key phases—such as animal studies, clinical trials, and other testing stages—and use Bidirectional Encoder Representations from Transformers (BERT) to perform Named Entity Recognition (NER), identifying key regulatory terms and entities. By converting these texts into numerical representations and segmenting them by phase, country, and year, we compare jurisdictional requirements and assess their alignment. Additionally, we apply Latent Dirichlet Allocation (LDA) for theme analysis to observe changes in regulatory focus over time, reflecting evolving priorities and challenges. Our analysis reveals notable semantic similarities and differences between countries and phases. For instance, the closest alignment in animal study regulations is between China and the USA, with a mean cosine distance of 0.33. These findings highlight the computational potential in regulatory science, offering valuable insights for researchers, policymakers, and industry professionals. 
\end{abstract}


\section{Introduction} 
Global regulatory authorities for medical products, such as the Food and Drug Administration's (FDA), prioritize two pillars when reviewing medical device dossiers to determine market eligibility: safety and efficacy \cite{darrow2021fda}. Safety ensures that medical products do not pose harm to patients, while efficacy confirms that the products achieve their intended indications when used. Although efficiency—the speed at which valuable devices reach patients—is also valued, it tends to receive comparatively less emphasis. Ensuring that high-quality devices reach patients worldwide remains critically important for those in need. However, products that successfully enter several markets often face significant challenges in accessing others, reflecting a disparity not always evident at first glance.

In 2018, Fractyl Laboratories' Revita DMR device, developed to treat insulin-resistant type 2 diabetes, received the FDA's Breakthrough Device Designation, intended to accelerate its development and review process~\cite{fractyl2023revita}. However, despite the expedited pathway in the US, the device faced delays in launching in other markets due to differing regulatory requirements. These varying regulations across regions significantly impacted the timeline for global availability. After substantial investment in research and development, the regulatory approval process can become a bottleneck, delaying market entry, and limiting patient access. This highlights the critical need to understand and navigate the diverse regulatory landscapes to ensure that innovative treatments reach patients in a timely manner.

The hip implant scandal in the EU has led to numerous recalls and reoperations for affected patients, highlighting serious safety concerns \cite{cohen2012fake}. The European regulatory system for medical implants has come under intense scrutiny following several high-profile device failures and recalls, resulting in additional patient investigations and surgeries. Unlike the centralized authority of the US FDA, Europe relies on more than 30 decentralized notified bodies to assess the safety and efficacy of medical devices \cite{cohen2012fake}. Investigations have revealed significant variations in regulatory practices, including differences in adopted standards, premarket evidence requirements, post-market surveillance, and factory audits. These inconsistencies create opportunities for manufacturers to take advantage of the approval process \cite{cohen2012fake}, demonstrating how regulatory frameworks can directly affect both the market entry and the safety of medical devices.

The International Medical Device Regulators Forum (IMDRF) seeks to accelerate the harmonization and convergence of global medical device regulations~\cite{imagawa2018regulatory}. While the IMDRF has made notable progress in aligning regulatory frameworks, the rules remain complex and continue to evolve rapidly, making it difficult for stakeholders to manually track and understand the differences across regions~\cite{han2024more}. These disparities in standards and requirements create a challenging regulatory landscape for manufacturers, often acting as barriers that delay or block patient access to medical products. As a result, patients may face delays or limited availability of innovative, potentially life-saving treatments. Bridging these regulatory gaps is crucial for streamlining approval processes, reducing time to market, and ensuring patients worldwide benefit from advancements in medical technology~\cite{tamura2014multiregional, engberg2015regulation}. 

To address this issue, our research seeks to answer the following question: How do regulatory requirements for medical devices differ across countries, and what are the practical implications of these differences for manufacturers? Specifically, how can manufacturers efficiently extract relevant information from the vast and continuously growing body of regulations and make informed decisions? Given the IMDRF's focus on convergence, how does each country align its regulatory framework across different phases, such as animal studies, clinical trials, and testing? Given the impracticality of manually reviewing all regulatory documents, we explore how natural language processing methods can assist in this process. Using real-world data, this study highlights the current state of regulatory harmonization and the challenges manufacturers face. By utilising natural language processing (NLP) methods, we are able to quantify and compare regulatory texts, offering a novel approach to measure of harmonization across regions.

\section{Background and Motivation}

Regulatory frameworks play a crucial role in ensuring the safety, efficacy, and quality of medical devices, governing every phase from pre-clinical testing to clinical trials and post-market surveillance. Regulatory affairs teams act as liaisons between companies and regulatory bodies, ensuring that products meet all legal and safety standards to facilitate market entry \cite{patil2023artificial}. However, these frameworks vary significantly across regions due to differences in national regulatory environments, creating a complex landscape for stakeholders to navigate \cite{fink2023comparison, manu2022review, ratna2018requirements, jyothi2013regulations, kramer2020regulation}. This complexity demands sophisticated analytical tools to effectively understand and manage the evolving regulations, as the challenge of regulatory complexity has been a topic of increasing scholarly interest~\cite{parker2012measuring, han2024more}. For manufacturers, researchers, regulatory professionals, and policymakers, keeping up with these changes is essential for compliance and fostering innovation.

Research has highlighted differences across various aspects of regulatory systems, such as quality systems, post-marketing requirements, and premarket approval processes. At the regulatory authority level, the U.S. FDA operates a centralized, adaptive process that allows rapid adjustments in device classifications based on new market data, which facilitates the timely introduction of innovative technologies. In contrast, the EU's Medical Device Regulation (MDR) relies on a more decentralized approach, using notified bodies to assess high-risk devices. While this ensures rigorous safety standards, it can lead to variability in enforcement across member states. The regulatory systems of both the U.S. and the EU are well-established for clinical trials and animal studies, and China’s regulatory landscape is rapidly evolving as it strives to align with international norms \cite{dhiman2021comparative}.

From a procedural standpoint, gaining regulatory approval varies significantly across regions. In Canada, manufacturers must submit an Investigational Testing Authorization (ITA) to Health Canada for Class II, III, and IV devices. In the EU, manufacturers must notify the competent authorities of member states before conducting clinical investigations. In the U.S., approval processes differ based on whether a device is classified as Significant Risk (SR) or Non-Significant Risk (NSR), which determines if an Investigational Device Exemption (IDE) or Institutional Review Board (IRB) approval is required \cite{chen2018comparative}. On a product-specific level, even identical medical devices, such as contact lenses, face different regulatory requirements in the U.S., EU, and other countries, leading to varying compliance obligations \cite{zaki2019review}. These discrepancies illustrate the challenges manufacturers face when navigating global regulatory landscapes, impacting the timely availability of medical devices worldwide.

\subsection{Regulatory Challenges}

The global launch of medical devices is often hindered by the diverse and complex regulatory requirements across different regions. For example, even after successfully passing the U.S. FDA’s rigorous review process, a medical device must still undergo additional approval processes to enter other markets, such as the EU or China, due to differing regulatory standards \cite{ramakrishna2015medical, han2024regulatory}. These regional differences necessitate distinct approval procedures for manufacturers, complicating the regulatory process. In addition to initial market entry, manufacturers must also manage ongoing regulatory obligations such as device change control, complaint handling, vigilance, and recall activities during the product lifecycle in each region where their devices are registered \cite{badnjevic2022post}.

\subsection{Efforts at Harmonization}

The International Medical Device Regulators Forum (IMDRF) has played a significant role in the harmonization of global medical device regulations by building on the foundation set by the Global Harmonization Task Force (GHTF). Since its establishment in 2011, IMDRF has focused on aligning regulatory approaches and improving cooperation between regulatory bodies from various countries, including the U.S., EU, China, Australia, Japan, and others \cite{tamura2014multiregional}. The International Medical Device Regulators Forum (IMDRF) has made significant efforts to harmonize various aspects of medical device regulation, including animal studies, while IMDRF's primary focus has been on clinical evidence, quality management systems.

IMDRF has published key documents such as the "Essential Principles of Safety and Performance of Medical Devices" and established working groups to address specific areas. The working focus includes harmonizing clinical evidence requirements and premarket submissions, which influence how animal studies and clinical trials are conducted across jurisdictions. Specific guidance related to clinical evidence evaluation, such as clinical investigations, has been designed to ensure that trial data from one country can be accepted by others, promoting international data sharing and reducing redundancy in testing. Some of the key IMDRF publications related to clinical evaluations and safety include frameworks the Clinical evaluation guidance (IMDRF/PMD WG/N56) \cite{N562022} and Post-market adverse event reporting guidance (IMDRF/NCAR WG/N14) \cite{N142023} and so on.

 Despite these efforts, challenges remain due to the diversity of regulatory institutions and the influence of various stakeholders, including industry, physicians, and scientific advisors \cite{altenstetter2012medical}. A deep understanding of these regulations and strategic planning is essential for manufacturers to successfully launch products in multiple regions.

\subsection{The Potential Use Computational Approaches in Regulatory Analysis}

Machine learning involves training a computational model to perform tasks by leveraging data. Broadly, these tasks fall into two main categories: prediction and data exploration \cite{nay2018natural, xiong2020dynamic}. Prediction relies on a specific branch of machine learning known as supervised learning, where data examples—like medical device regulations—include pairs of (i) predictor variables (such as a device’s manufacturer or classification) and (ii) an outcome variable (for instance, whether the device achieved regulatory approval). With this structure, a model can learn to process new observations of the same predictor variables (e.g., a new device's manufacturer and classification) and predict its likely outcome (approval status).

Data exploration is often achieved using techniques known as unsupervised learning. Unlike supervised learning, unsupervised learning deals with data that only include observed variables without a designated outcome variable to predict. This allows for a larger pool of data since most datasets lack explicit outcome labels. For instance, there is a wealth of raw medical data available from various regulatory submissions. In the context of medical device regulation, a task well-suited for unsupervised learning might involve identifying other devices with similar regulatory profiles or technical features to a given device.

Complementing these machine learning methods, Natural Language Processing (NLP) has become increasingly valuable for enhancing the efficiency and accuracy of regulatory analysis. As a branch of artificial intelligence, NLP enables the automated understanding and extraction of information from regulatory documents, which is particularly useful when handling large volumes of text. Recent advancements in NLP have made it possible to automate the extraction of critical insights from massive collections of regulatory documents, allowing us to detect intricate patterns and trends that might be overlooked in a manual review \cite{jallan2019application}. Together, these computational approaches provide a more comprehensive toolkit for managing and analyzing regulatory data in a more efficient and insightful manner. As the regulatory landscape and medical devices evolve rapidly, maintaining harmonization across regions becomes more challenging, highlighting the growing importance of AI-driven tools \cite{seyhan2019lost, zou2024dyna}.

For example, NLP techniques like BERT have been employed to analyze regulatory texts within the European Union, revealing that specific therapeutic areas align more closely with European Medicines Agency (EMA) guidelines than others \cite{bergman2023full}. In addition, unsupervised learning methods like Latent Dirichlet Allocation (LDA) have proven effective in identifying and categorizing regulatory themes, such as safety, efficacy, and compliance, from large datasets \cite{blei2003latent}. These methods significantly reduce the manual effort involved in understanding complex regulatory frameworks. Machine learning approaches have also been applied to automatically classify regulatory submissions, improving efficiency and accuracy in the submission and review process \cite{ceross2021machine}. NLP-based analysis of FDA approval letters, for instance, has helped to identify recurring themes and areas needing improvement, thus enhancing the regulatory review process \cite{gurulingappa2012extraction}. Similarly, NLP tools have been used to extract key data from clinical trial registries, facilitating better analysis of trial designs and outcomes \cite{yim2016natural}.

Furthermore, temporal analysis of regulatory documents, enabled by computational tools, offers insights into how regulatory priorities shift over time. This is particularly valuable in response to major events, such as the COVID-19 pandemic, during which regulatory agencies worldwide accelerated the approval processes for diagnostic and therapeutic devices, demonstrating remarkable flexibility \cite{beninger2020covid}. Time-series analysis has also tracked growing emphasis on areas like cybersecurity and data protection in medical device regulations over the past decade \cite{haque2020illuminating}. Such computational analyses not only help stakeholders stay informed about changing regulatory landscapes but also allow them to better anticipate future trends, enabling proactive decision-making and strategic planning \cite{xiong2019analysis}.

\subsection{Contributions}

Our research tackles the challenges stemming from the lack of harmonization in medical device regulations, which often leads to delays in product approvals and can pose safety risks. Regulatory guidelines are typically communicated in written form to various stakeholders, including regulatory bodies, manufacturers, Contract Research Organizations (CROs), and patients. Traditionally, researchers have manually compared regulatory frameworks and summarized their findings for the scientific community \cite{kramer2014ensuring}. However, the rapid evolution of these regulations makes manual comparisons inefficient and prone to errors. By applying NLP techniques, we aim to transform regulatory texts into numerical embeddings, laying the groundwork for a more efficient, accurate, and scalable approach to analyzing global regulatory frameworks. This method holds the potential to reduce barriers to international market entry for medical devices, enabling faster and safer access to innovations. The yearly results of the LDA analysis across different countries also reveal shifts in regulatory themes within the medical device sector.

\section{Methodology} \label{Section: model}
This study utilizes BERT embeddings and Latent Dirichlet allocation (LDA) \cite{xie2020monolingual} to analyze a comprehensive dataset of 664 regulations—over 200 million tokens—primarily comprising technical regulations and guidelines from the USA, EU, and China issued over the past decade. We categorize these regulations into distinct phases: animal studies, clinical trials, and other testing stages, allowing for a comparative assessment of each country's requirements and the degree of alignment among them. To achieve this, we employ specific keywords to filter and group phase-specific sentences, resulting in tailored corpora for each regulatory phase. We then leverage BERTSum, a summarization model built upon BERT (Bidirectional Encoder Representations from Transformers) \cite{liu2019text}, to gain insights into the underlying reasons for observed regulatory differences and their implications for manufacturers. BERTSum is particularly useful for extracting key points from extensive regulatory documents, providing a concise understanding of each region's regulatory focus.

\subsection{Data Collection and Pre-processing}

All regulations and guidelines were sourced from official government websites and databases specific to the USA FDA \cite{FDA2024}, EU EMA \cite{MDCG2024}, and China NMPA \cite{NMPA2024}. These documents, manually downloaded, pertain to medical devices and, in some instances, combination products, which include both devices and pharmaceutical products. The dataset encompasses laws, guidelines, and policies relevant to medical devices, available in English as well as the native languages of the respective regions. The collection period spanned from Jan 2013 to May 2024. Only documents directly providing guidance on medical devices were included, while documents such as meeting minutes and results of fly surveillance were excluded. All documents were stored in a standardized format for further analysis. The text extraction and analysis were conducted using Python (e.g., Python 3.9), with libraries such as NLTK \cite{bird2009natural} for natural language processing and SciPy \cite{virtanen2020scipy} for statistical analysis. These libraries facilitated efficient data processing, tokenization, and semantic similarity calculations. The twenty regulations are listed in Table~\ref{tab:regulations}.

\begin{table}[h!]
\centering
\begin{tabular}{|p{1cm}|p{10cm}|p{2.9cm}|}
\hline
\textbf{Year} & \textbf{Title} & \textbf{Authority} \\ \hline
2024 & Guiding Principles for the Registration and Review of Laparoscopic Surgery Systems, Animal Testing, Decision-making, Judgment and Requirements & China NMPA \\ \hline
2023 & Guidelines for the Registration Review of Medical Molecular Sieve Oxygen Concentrators & China NMPA \\ \hline
2022 & Single-Use Cervical Balloon Dilation Catheter Registration Review & China NMPA \\ \hline
2022 & Provisions for Supervision and Administration of Medical Device Manufacturing & China NMPA \\ \hline
2021 & Medical diagnostic X-ray equipment for pediatric applications & China NMPA \\ \hline
2020 & Hernia Repair Mesh Clinical Trial Guidelines & China NMPA \\ \hline
2014 & Renewal of EC Design-Examination and Type-Examination certificates & EU EMA\\ \hline
2017 & Designation and notification of conformity assessment bodies & EU EMA\\ \hline
2019 & GUIDANCE NOTES FOR MANUFACTURERS of class I medical devices & EU EMA \\ \hline
2020 & Post-market clinical follow-up (PMCF) Plan Template & EU EMA\\ \hline
2022 & MDCG Position Paper Notice to manufacturers to ensure timely compliance with MDR and IVDR requirements & EU EMA\\ \hline
2024 & Guidance on content of the Investigator’s Brochure for clinical investigations of medical devices & EU EMA\\ \hline
2013 & Establishing the Performance Characteristics of In Vitro Diagnostic Devices for the Detection of Antibodies to Borrelia burgdorferi & USA FDA \\ \hline
2014 & Infusion Pumps Total Product Life Cycle & USA FDA \\ \hline
2016 & Assessment of Radiofrequency-Induced Heating in the Magnetic Resonance (MR) Environment for Multi-Configuration Passive Medical Devices & USA FDA \\ \hline
2017 & FDA Categorization of Investigational Device Exemption (IDE) Devices to Assist the Centers for Medicare and Medicaid Services (CMS) with Coverage Decisions & USA FDA \\ \hline
2019 & Recommended Content and Format of Non-Clinical Bench Performance Testing Information in Premarket Submissions & USA FDA \\ \hline
2021 & Safer Technologies Program for medical devices & USA FDA \\ \hline
2023 & Best Practices for Selecting a Predicate device to support a premarket notification submission & USA FDA \\ \hline
2024 & Perform servicing and Remanufacturing & USA FDA \\ \hline
\end{tabular}
\caption{Partial Overview of Medical Device Regulations and Guidelines Issued by Authorities (2013–2024)}
\label{tab:regulations}
\end{table}

All documents were converted to plain text and subsequently segmented, tokenized, and normalized. The Chinese regulations were translated into English and subsequently cross-checked through back-translation by two independent authors to ensure accuracy and consistency. The preprocessing of text involved several steps to clean and prepare the data for analysis. This included removing uninformative text such as boilerplate text, headers, footers, and repeated sections; breaking down the text into individual words or tokens; eliminating common stop words like "the," "and," and "in" that do not contribute to meaningful analysis; and normalizing the text by converting it to lowercase and stemming words to their root form to ensure consistency in analysis, treating terms like "regulations" and "regulation" as the same.

\begin{figure}[t]
   \centering
\includegraphics[scale=0.3]{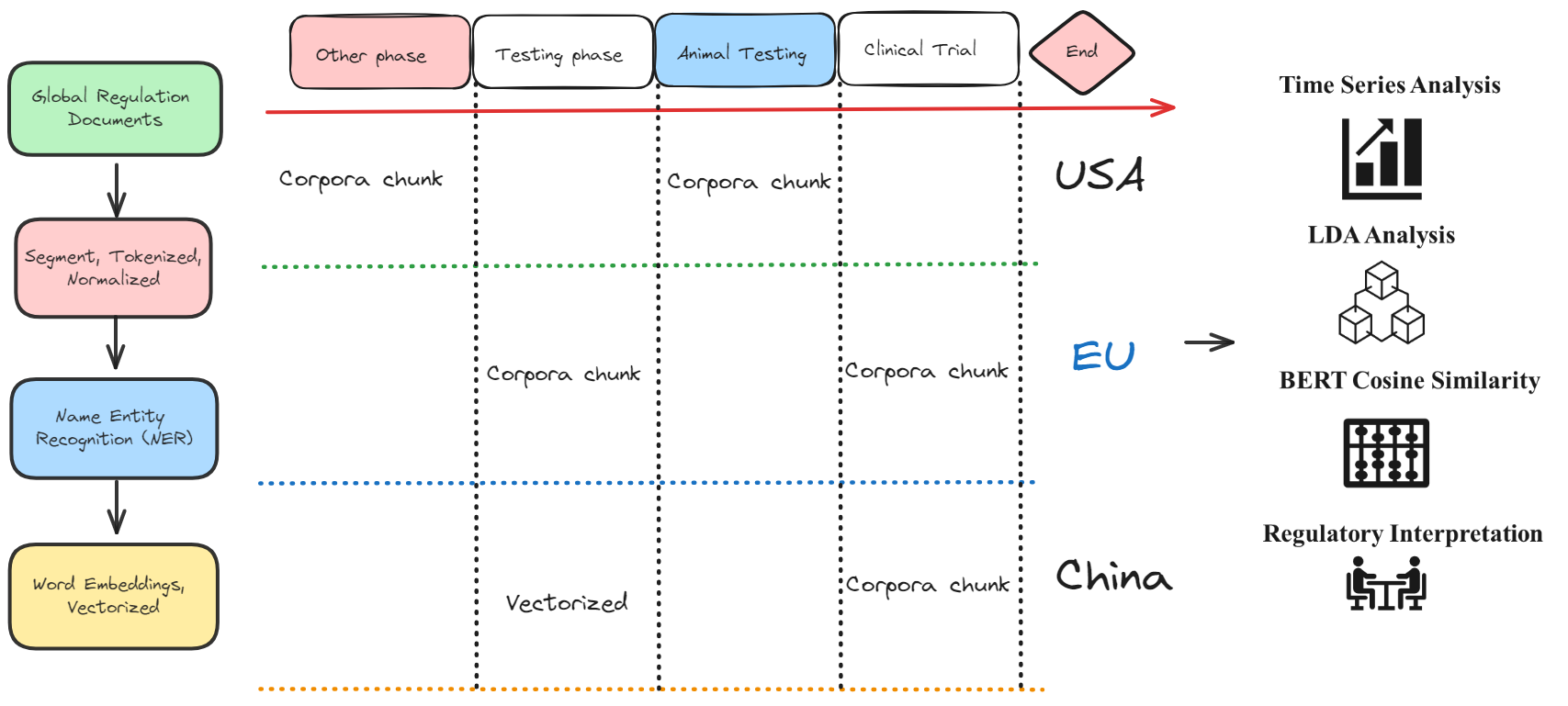}
    \caption{The Workflow of Text Cleaning and Preprocessing}
   \label{fig: 1}
\end{figure}

\subsection{Name Entity Recognition}

Named Entity Recognition (NER) is a sub-task of information extraction that aims to locate and classify named entities from unstructured text into predefined categories such as names of persons, organizations, locations, expressions of time, quantities, monetary values, and percentages. NER is a fundamental component in NLP applications, including information retrieval, question answering, and machine translation. This process can be achieved through machine learning approaches and deep learning models.

In our study, NER is specifically applied to extract relevant biomedical terms from text, recognizing the unique nature of the biomedical field. We utilize SciSpacy, a Python package that provides language models optimized for processing biomedical, scientific, and clinical text. SciSpacy enables us to effectively identify entities in this specialized context~\cite{scispacy2024}. Given that biomedical concepts are central to medical device regulations, applying NER allows us to capture a wide array of relevant biomedical names.

We applied NER to the corpora for each country, analyzing data on an annual basis. Additionally, SpaCy~\cite{spaCy2024} was used for text preprocessing and NER analysis, while Matplotlib and Seaborn were employed for data visualization.

\subsection{Subgroup Annotation}

\begin{figure}[htbp!]
   \centering
   \includegraphics[scale=0.44]{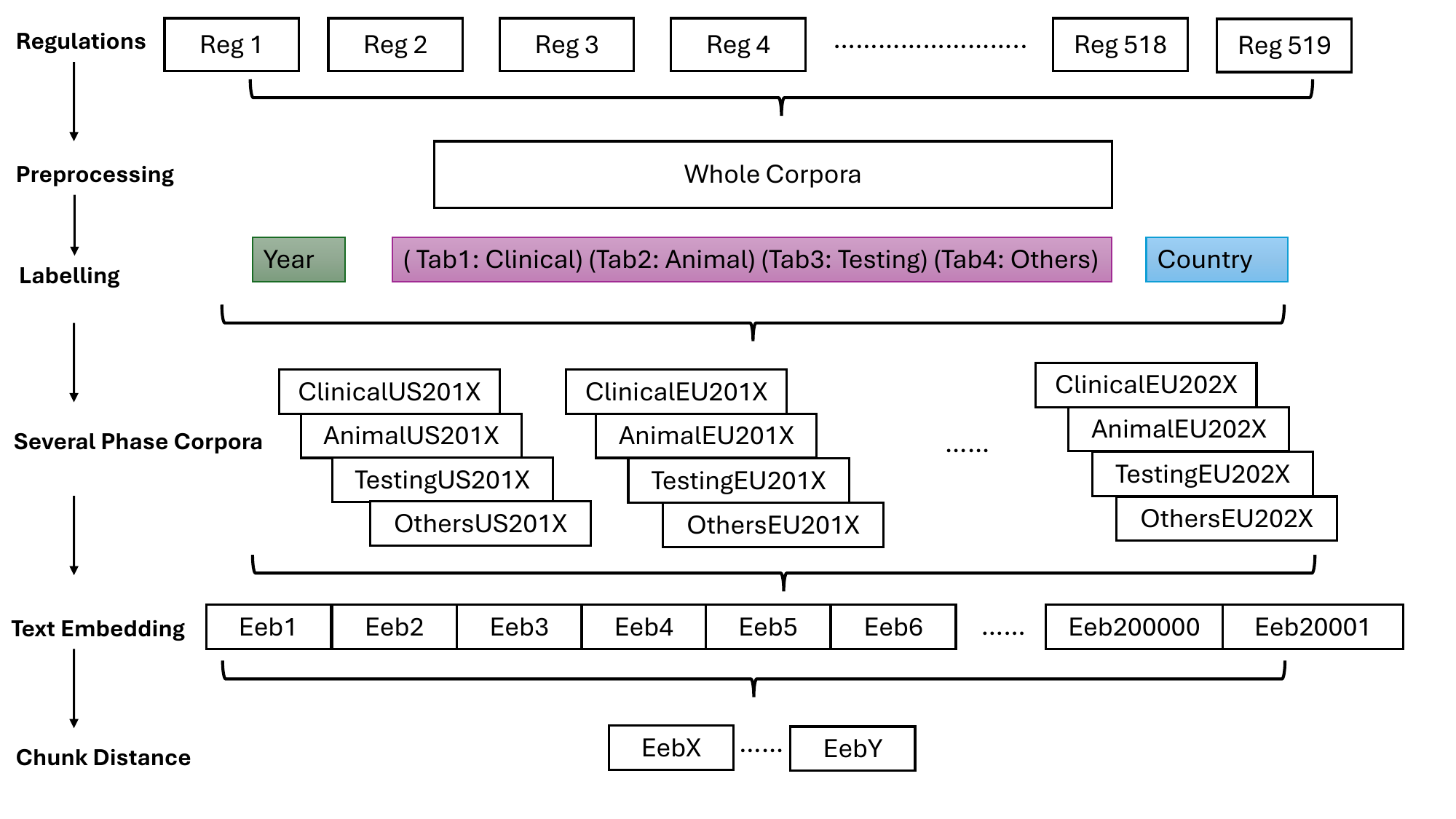}
   \caption{Segmentation of Regulations: Labelling, Chunking, Embedding, and Distance Calculation}
   \label{fig: BERT}
\end{figure}

\begin{figure}[t]
   \centering
\includegraphics[scale=0.65]{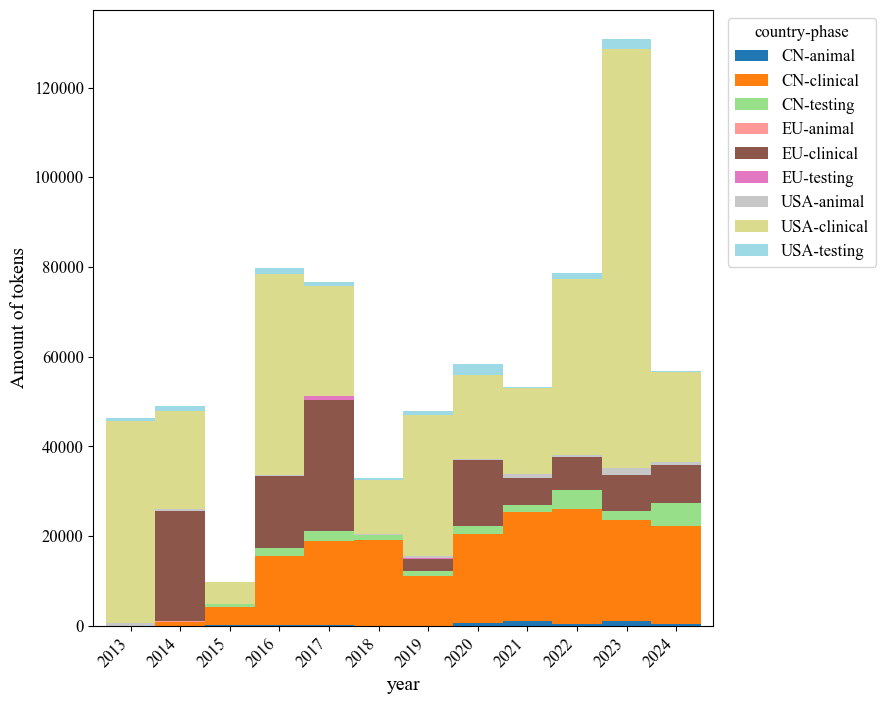}
    \caption{Quantitative corpora amount of different chunks}
   \label{fig: co}
\end{figure}

The development of medical devices is a complex and highly regulated process requiring rigorous testing and evaluation to ensure safety and efficacy before approval. The regulatory landscape aims to protect public health by enforcing stringent standards throughout the device lifecycle, from research and development to post-market monitoring. In this study, we categorize regulatory documents into four main phases relevant to the medical device lifecycle: preclinical testing (animal studies), clinical trials, testing phase, and other phases. This classification is informed by industry knowledge and reflects the characteristic stages of medical device development, highlighting the significance of each phase within the regulatory framework. In Figure \ref{fig: co}, we visualize each chunked corpus. To enhance clarity, we excluded other corpora due to their volume, allowing for a clearer view of the animal, testing, and clinical corpora in this figure.

\textbf{Animal Studies}: The phase of animal studies, focuses on preclinical evaluations of safety and biological effects~\cite{cheng2022regulatory}. These studies gather preliminary data on the interactions between the device and living tissues, assessing factors such as toxicity, biocompatibility, and potential side effects. Insights gained from animal testing are crucial, as they cannot be ethically or feasibly obtained through in vitro (lab-based) tests alone. We filtered relevant text from the corpus using keywords such as ``animal testing,'' ``animal models,'' ``animal toxicity tests,'' ``animal ethics,'' and ``animal.'' The extracted sentences were compiled into a corpus for subsequent analysis.

\textbf{Clinical Trials}: The clinical phase involves testing the medical device on human subjects to assess its safety and effectiveness in real-world settings \cite{petryna2009experiments}.  This phase encompasses various stages of clinical trials, ranging from small pilot studies to large randomized controlled trials. Human trials are indispensable for understanding device performance and identifying potential risks. To ensure comprehensive coverage of pertinent documents, we employed keywords including ``clinical,'' ``clinical research,'' ``patients,'' ``clinical evaluation,'' ``randomized controlled trial,'' ``safety assessment,'' ``efficacy,'' ``pharmacodynamics,'' ``side effects,'' ``subjects,'' and ``human trials.'' These trials are rigorously regulated to protect participant rights while generating robust data for regulatory approval.

\textbf{Testing Phase}: This phase includes all activities related to lab testing, quality control, testing methods, and procedures conducted before the registration of medical devices. It also involves ongoing monitoring to ensure that devices consistently meet safety and performance standards after the market release \cite{panteghini2023redesigning}. We captured relevant regulatory documents using keywords such as ``testing methods,'' ``quality control,'' ``analytical methods,'' ``laboratory testing,'' ``test standards,'' ``testing equipment,'' ``calibration,'' ``sample collection,'' ``data analysis,'' ``instrument validation,'' ``standard operating procedures,'' and ``testing.'' These documents guide manufacturers and regulatory bodies in maintaining high standards of device quality and patient safety.

\textbf{Other Phases}: Documents that do not fit into the predefined phases are classified under ``Other Phases.'' This category includes various regulatory and compliance-related activities that support the overall lifecycle of medical devices but do not strictly fall into animal testing, clinical trials, or post-market testing. Including an ``Other Phases'' category ensures that no pertinent regulatory information is overlooked, allowing for a comprehensive analysis that captures broader regulatory compliance, risk management, labeling, and other essential activities contributing to the safe and effective use of medical devices.

\subsection{Utilizing BERT embedding for cosine similarity for Regulatory Document Analysis}

As illustrated in Figure \ref{fig: BERT}, the entire set of regulations was pre-processed into a comprehensive database referred to as the whole corpus. Each document within this database was labeled with various keywords to filter and segregate the text into distinct phases of regulation. Additionally, documents were tagged with country and year information, allowing the creation of several smaller, distinct corpora. For instance, a corpus labeled \texttt{Clinical\_CH2011} includes all clinical-related paragraphs and words from Chinese regulations in 2011.

BERT is a pre-trained transformer model specifically designed to understand the context of words in a sentence by considering the surrounding words \cite{zhang2020semantics}. We selected BERT as our embedding method because bag-of-words does not account for word order or context, and TF-IDF also overlooks the nuanced relationships between words by focusing solely on term frequency and document rarity \cite{sun2022text}. By applying BERT to segmented corpora, such as Clinical\_CH2011, we generated embeddings for each document. Subsequently, we assessed the similarity between these corpora using cosine similarity measures. Detailed statistical analyses on these similarity scores allowed us to identify patterns and draw meaningful conclusions about the relationships between different phases.

The categorization of regulatory documents into these distinct phases provides a structured approach to analyzing the complex landscape of medical device regulation. By focusing on the Animal Phase, Clinical Phase, Testing Phase, and Other Phases, we can thoroughly examine all relevant guidelines and requirements, facilitating a comprehensive understanding of the regulatory environment governing medical device development and post-market monitoring. As shown in Figures \ref{fig: 1} and \ref{fig: 2}, we visualized the corpus chunks for different phases in each country across various years.

\subsection{Utilizing LDA for Topic Modeling in Regulatory Document Analysis}

Following the embedding process, we utilized Latent Dirichlet Allocation (LDA)~\cite{blei2003latent} to uncover the underlying topics within the regulatory documents. LDA posits that each document is a mixture of several topics, with each word attributable to one of these topics. This probabilistic model helps us discover the hidden structure of the text by grouping words that frequently co-occur, thereby delineating distinct themes.

We applied LDA to the entire corpus to identify overarching topics relevant to medical device regulation. Additionally, we conducted a time series analysis, performing LDA on the corpora from each country for each year. This analysis enables us to examine how topics have evolved over time across different regions. Initially, the LDA model was trained using the preprocessed text data, and the number of topics was determined by evaluating the coherence score, which assesses the interpretability of the identified topics \cite{blei2003latent}. 

Each topic was carefully reviewed and labeled according to its most representative words. To facilitate better understanding, we categorized these topics under broader themes. Furthermore, visualization tools were employed to graphically represent the topics, enhancing the interpretability of the results \cite{mabey2018pyldavis}. This methodical approach ensures a robust analysis of the regulatory documents, providing valuable insights into the underlying thematic structures within the regulatory landscape.
\section{Results}
\label{sec:results}

We labeled all regulations by their respective year and country, then visualized the trend in Figure \ref{fig: 1}. The line graph illustrates changes in the number of regulatory documents issued by China (CN), the European Union (EU), and the United States (USA) from 2013 to 2024. Each line represents the trend of document issuance for one of these regions. The graph reveals an upward trend for every region, indicating that the number of regulations published annually has increased across all three jurisdictions.

\subsection{Increasing Trends in Regulatory Issuance Across the USA, EU, and China}
From data shown in Figure \ref{fig: 1}, we found the United States exhibits a volatile trend in regulatory document issuance. The most pronounced increase occurs between 2020 and 2022, reaching the highest peak among the three regions. Conversely, China demonstrates a fluctuating trend characterized by a general increase over the years. Starting with a moderate count in 2014, the issuance in China displays noticeable peaks and troughs, with significant growth observed after 2020, culminating in a peak around 2022. This suggests intensified regulatory activity in recent years, potentially driven by evolving industry standards and regulatory requirements. In contrast, the trend for the European Union is relatively stable, with only minor fluctuations.

The trend of increasing regulatory guidelines issued by the FDA marks a significant shift in the agency's approach to managing medical device approval and oversight. Historically, the number of guidelines remained relatively constant; however, since the mid-2010s, there has been a notable surge \cite{daizadeh2021historically}. This surge can be partly attributed to the COVID-19 pandemic and the proliferation of digital health technologies. During this period, authorities issued numerous new guidelines to ensure the safety and efficacy of emerging technologies and emergency medical devices. The increase is not limited to COVID-19-related emergency authorizations but also encompasses regulations for digital health devices and other medical innovations \cite{daizadeh2021historically}.

This rapid increase reflects the FDA's efforts to address public health emergencies and adapt to the evolving landscape of digital health technologies. As technology continues to advance and medical devices evolve, we expect that global authorities will continue to expand their issuance of guidance to address the challenges and opportunities presented by these new technologies.

\begin{figure}[t]
   \centering
    \includegraphics[scale=0.55]{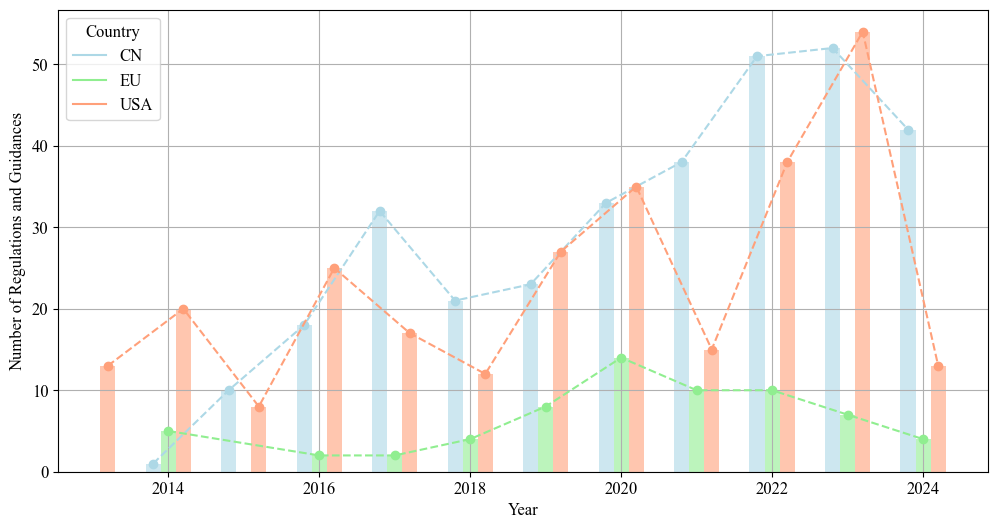}
    \caption{Quantitative analysis of regulation amount in past decade.}
   \label{fig: 1}
\end{figure}

\subsection{Regulatory Focus in Medical Devices: Insights from Named Entity Recognition Analysis}

The analysis of the top three Named Entity Recognition (NER) entities over time in the medical device sector for China, the European Union (EU), and the United States (USA) from 2013 to 2024 reveals significant trends and patterns in regulatory focus across these regions. For the frequency of various types of entities extracted from the analyzed text corpus, there are totally 270,257 entities for all countries regulation documents.

Our analysis of Named Entity Recognition (NER) in the medical device sector reveals that the most frequently identified entity types are CARDINAL and ORG, comprising 35.7\% (n = 96,531) and 36.4\% (n = 98,481) of the total entities recognized, respectively. Following these are DATE entities, accounting for 7.4\% (n = 19,880), and PERSON and LAW entities, which constitute 5.3\% (n = 14,200) and 4.1\% (n = 11,201), respectively. Geographic and product-related entities, represented by GPE and PRODUCT, make up smaller proportions at 3.1\% (n = 8,290) and 1.7\% (n = 4,520), respectively. Other entity types, such as NORP (0.96\%), ORDINAL (0.70\%), and PERCENT (0.51\%), each constitute around 1\% or lower. This distribution highlights the diverse range of entity types present in regulatory and technical texts related to the medical device field, with a notable emphasis on numerical and organizational references.

\begin{figure}[t]
   \centering
    \includegraphics[scale=0.5]{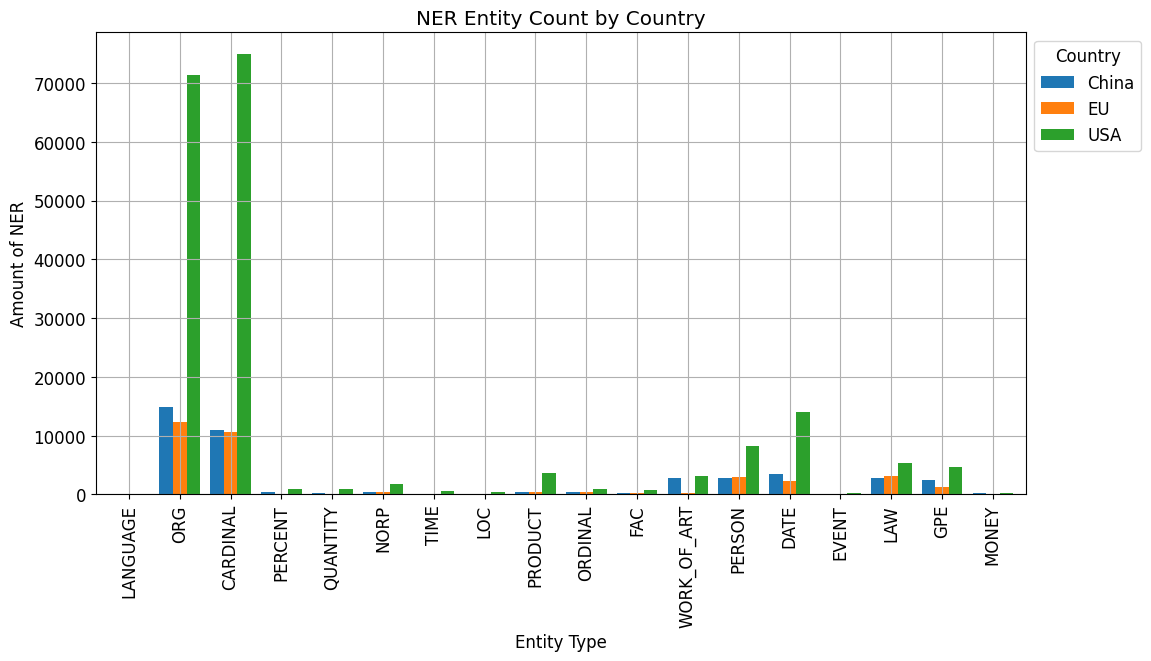}
    \caption{Quantitative analysis of regulation amount in past decade.}
   \label{fig: 1}
\end{figure}

This distribution emphasizes the strong presence of numerical values and organizational references, alongside temporal, personal, and legal elements within the analyzed corpus. Less frequent entities such as NORP (nationalities or religious/political groups), ORDINAL (ordinal numbers), PERCENT (percentages), FAC (facilities), QUANTITY (measurements), LOC (locations), TIME (specific times), MONEY (monetary values), EVENT (named events), and LANGUAGE (languages) further illustrate the varied nature of the text. The high frequency of CARDINAL and ORG entities indicates that the corpus contains substantial numerical data and organizational references, while the presence of DATE, PERSON, and LAW entities highlights a focus on temporal information, personal names, and legal content. This distribution provides valuable insights into the thematic emphasis and contextual framework of the content.

\begin{figure}[t]
   \centering
\includegraphics[scale=0.42]{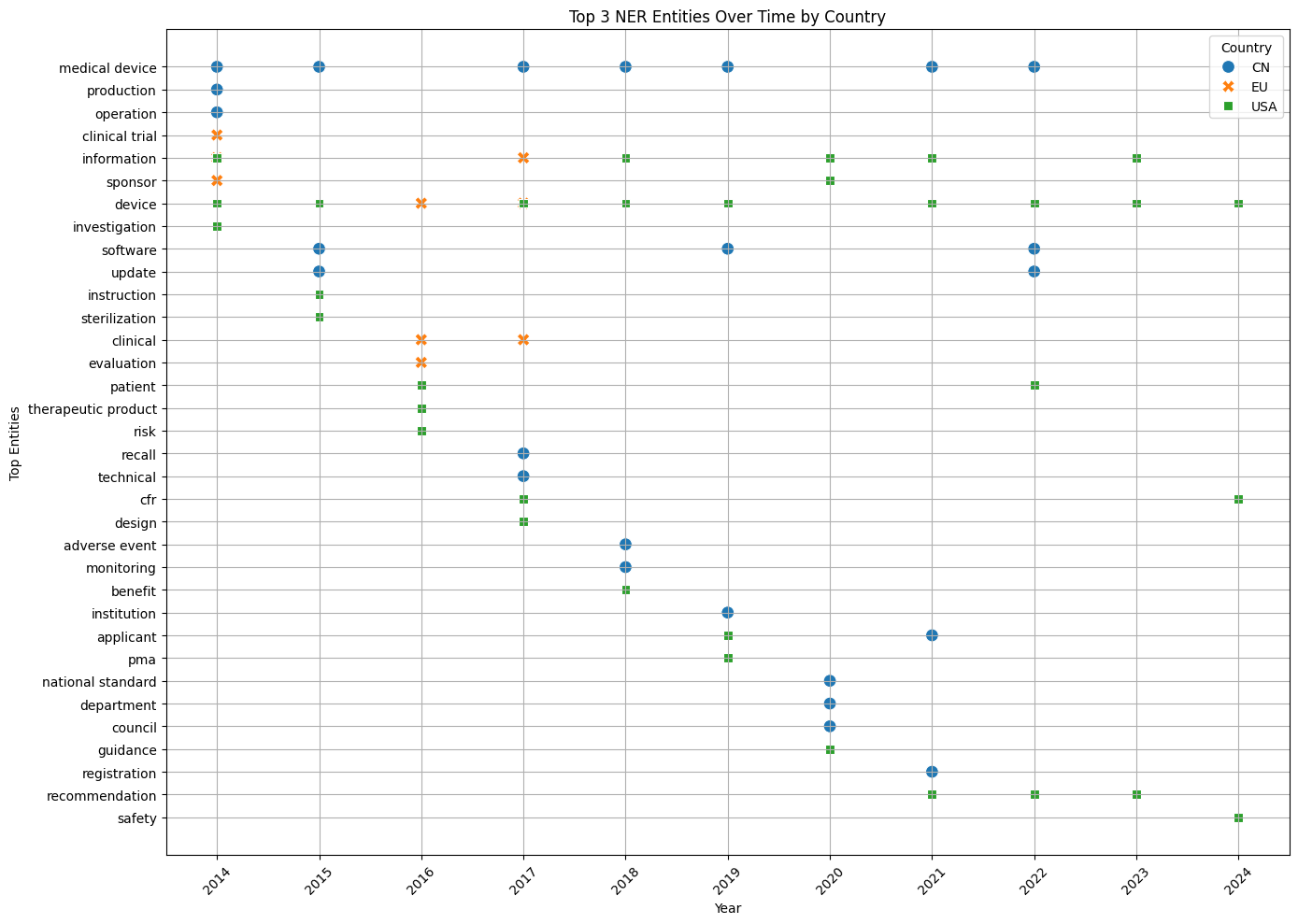}
    \caption{Top 3 NER Corpora by Year and Country}
   \label{fig: 2}
\end{figure}

From Figure~\ref{fig: 2}, the NER results indicate that in the USA, there was an early focus on "technical" terms and "CFR" (Code of Federal Regulations), which suggests strong regulatory activity around technical standards and compliance from 2014 to 2016. A noticeable increase in terms like "software" and "update" reflects the growing importance of digital health technologies and the need for updated regulatory frameworks during 2017-2019, as the FDA issued several guidance documents that year. This trend is consistent with observations made by other researchers~\cite{bhavnani20172017, guo2020challenges}. In the period from 2020 to 2022, peaks in mentions of "medical device" and "safety" are likely driven by the COVID-19 pandemic and the urgent need for safe and effective medical devices.

The EU and the USA show a strong focus on "clinical trial" and "evaluation," particularly around 2016-2017, reflecting an emphasis on clinical trials and the evaluation processes crucial for medical device approvals. This trend aligns with the implementation of the EU Clinical Trials Regulation No. 536/2014, which aimed to streamline clinical trial processes and ensure participant safety~\cite{EUR2014}. Entities such as "sterilization" and "safety" are also prominent, highlighting the EU's emphasis on maintaining high hygiene standards and ensuring device safety during the 2018-2020 period.

In the USA, terms such as "risk," "recall," and "safety" emphasize the importance of risk management and proactive measures to address defective or unsafe devices. Furthermore, China's frequent mention of "national standard," "department," and "council" points to efforts to align medical device regulations with both national and international standards, ensuring consistency and regulatory compliance. Understanding these trends will help regulatory authorities and manufacturers navigate the evolving landscape of medical device regulations. For regulatory bodies, it highlights areas requiring continuous improvement and alignment with global standards. For manufacturers, it underscores the importance of adhering to clinical trial data, robust software development practices, and stringent risk management protocols.

\subsection{Topic modelling: Latent Dirichlet Allocation (LDA) Topics and Their Themes}

We applied Latent Dirichlet Allocation (LDA) to the entire corpus to uncover the major themes present in regulatory documents. Table \ref{tab: 5} summarizes the top keywords for each identified topic, providing insights into the thematic structure of the texts.

\begin{table*}[!ht]
\centering
\begin{tabular}{|p{7cm}|c|p{6cm}|}
\hline
\textbf{Top Words} & $\rightarrow$ & \textbf{Theme Inferred} \\ \hline
mdr, mdcg, annex, article, notified, authority, designating, group, pediatric, nb & $\rightarrow$& \textbf{Theme 0: Regulatory Framework and Compliance} \\ \hline
article, device, investigation, irb, consent, subjects, mdr, informed, asca, accreditation &$\rightarrow$ & \textbf{Theme 1: Clinical and Ethical Guidelines} \\ \hline
registration, guidelines, equipment, applicants, indicators, production, diagnostic, reagents, packaging, trial & $\rightarrow$& \textbf{Theme 2: Registration and Diagnostic Equipment} \\ \hline
software, functions, updates, algorithm, documentation, business, cybersecurity, hardware, environment, mobile & $\rightarrow$& \textbf{Theme 3: Software and Cybersecurity} \\ \hline
clinical, study, trial, sponsor, article, member, reviewer, sponsors, author, benefit & $\rightarrow$& \textbf{Theme 4: Clinical Trials and Sponsorship} \\ \hline
study, recommend, tissue, subjects, population, sample, statistical, tests, disease, protocol &$\rightarrow$ & \textbf{Theme 5: Research and Statistical Analysis} \\ \hline
market, premarket, study, model, predicate, recommend, market, software, submissions, modification &$\rightarrow$ & \textbf{Theme 6: Market Approval and Recommendations} \\ \hline
market, device, submissions, decision, approval, data, premarket, new, program, submitter &$\rightarrow$ & \textbf{Theme 7: Regulatory Decisions and Data} \\ \hline
combination, constituent, applicant, device, reporting, pma, premarket, manufacturing, drugs, program &$\rightarrow$ & \textbf{Theme 8: Combination Products and Manufacturing} \\ \hline
animal, recommend, biocompatibility, tissue, material, iso, coating, tests, chemical, testing &$\rightarrow$ & \textbf{Theme 9: Biocompatibility and Material Standards} \\ \hline
\end{tabular}
\caption{Top Words and Their Derived Topics}
\label{tab: 5}
\end{table*}

\begin{figure}[htbp]
    \centering
    \includegraphics[scale=0.29]{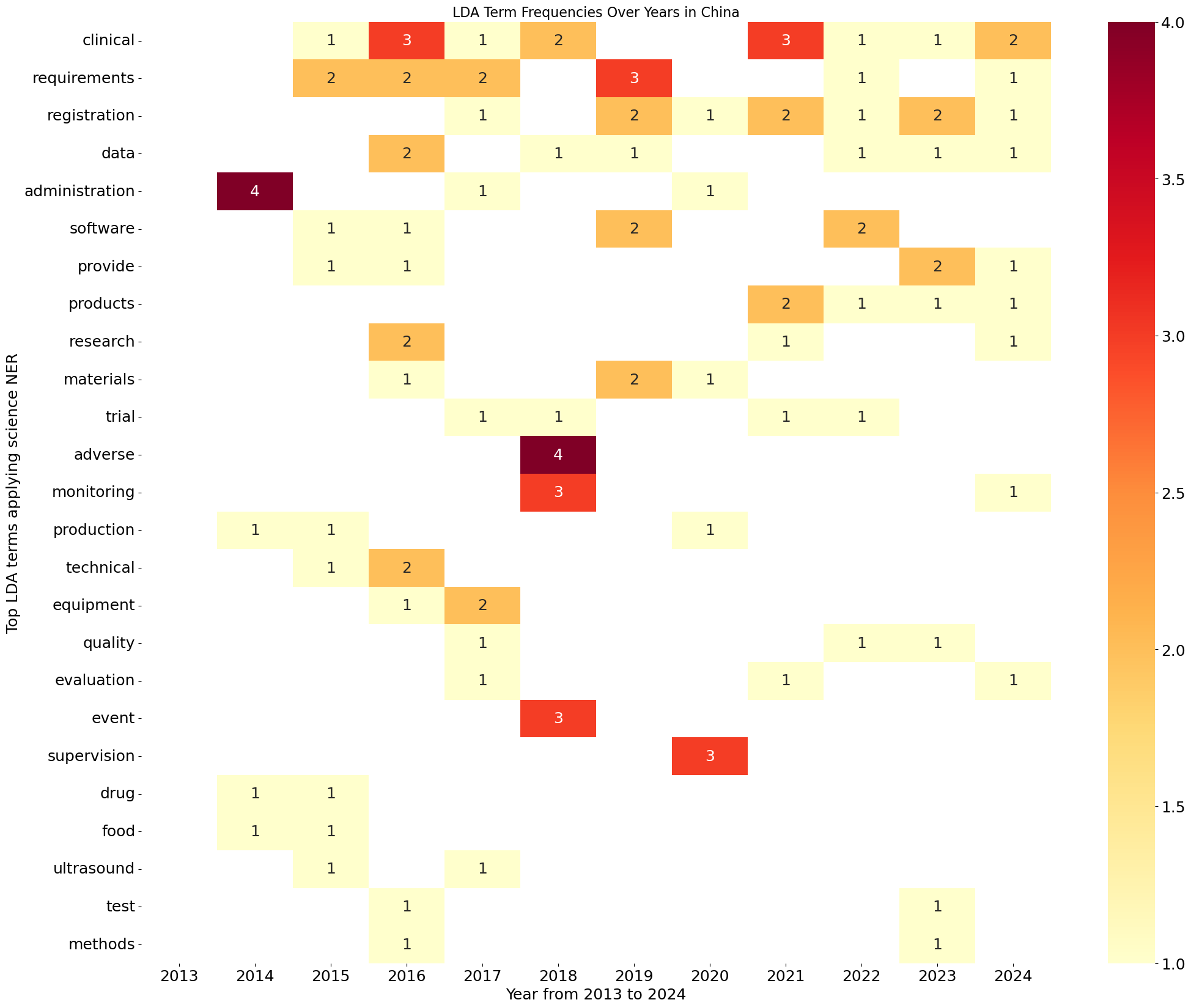}
    \caption{LDA of China corpora over years}
    \label{fig:cn2}
\end{figure}

\begin{figure}[htbp]
    \centering
    \includegraphics[scale=0.29]{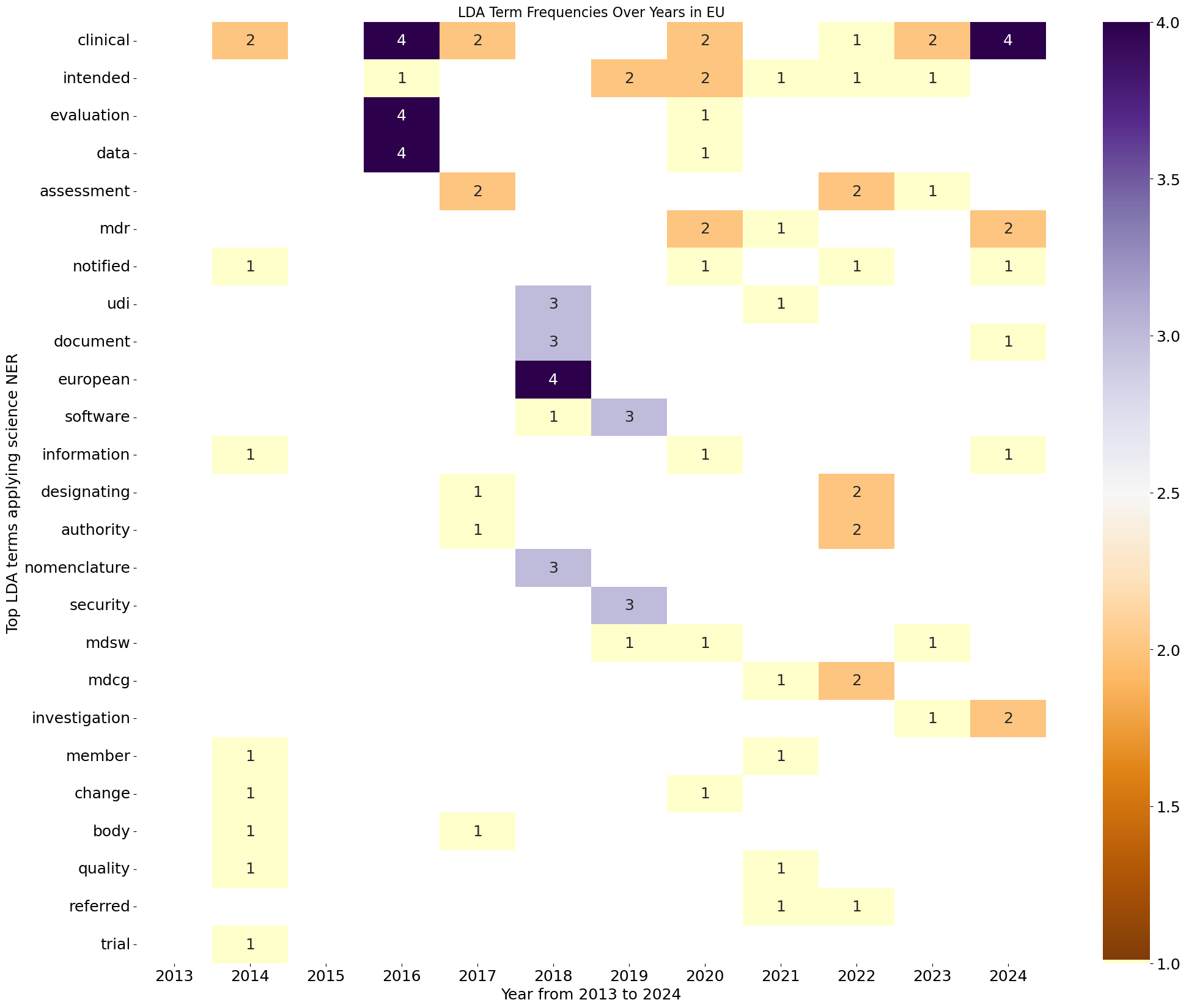}
    \caption{LDA of EU corpora over years}
    \label{fig:eu2}
\end{figure}

\begin{figure}[htbp]
    \centering
    \includegraphics[scale=0.29]{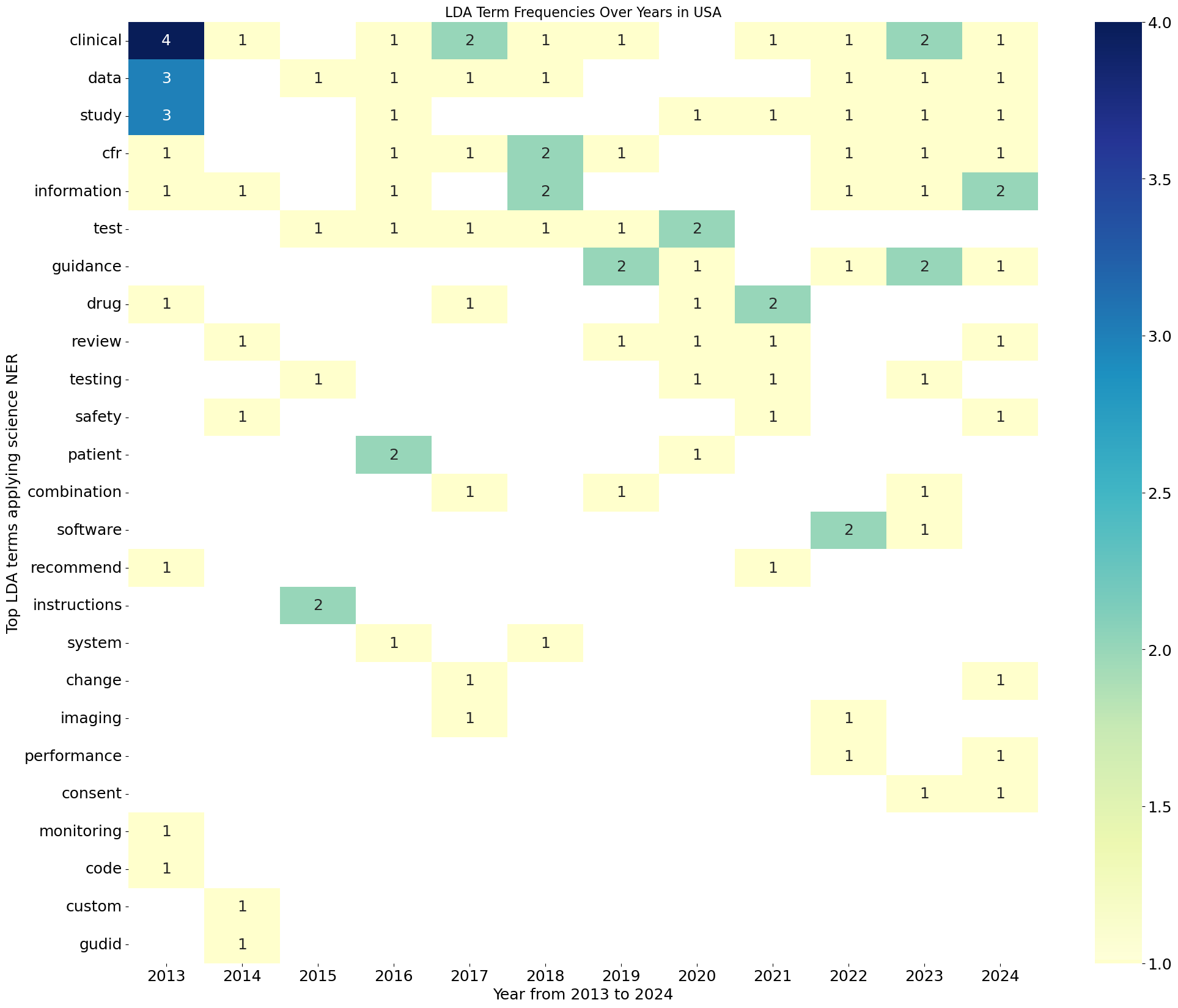}
    \caption{LDA of USA corpora over years}
    \label{fig:us2}
\end{figure}

We then applied LDA to each year's corpus from the different countries individually, generating a series of LDA topics. The results are visualized in Figures \ref{fig:cn2}, \ref{fig:eu2}, and \ref{fig:us2}, representing the term frequencies for China, the EU, and the USA over time. These visualizations reveal how the focus of each country changes from year to year, allowing us to observe shifts in thematic emphasis.

The numbers in each cell of the figures indicate the frequency of specific terms in the regulatory documents for that year, with cell colors corresponding to the term frequency. Darker colors represent higher frequencies, while lighter colors indicate lower frequencies. By tracking changes in color intensity and the associated numbers over time, we can identify trends in the topics emphasized within each country's regulatory framework. For example, the term "clinical" shows high frequency in the years 2013, 2015, 2016, 2019, and 2023, indicating a consistent focus on clinical topics during those periods. Similarly, terms like "evaluation" and "data" have high frequencies in 2015 and 2016, suggesting that these years saw heightened emphasis on regulatory evaluation and data-driven decisions. The term "software" shows moderate frequency in 2018 and 2019, reflecting a growing attention to software-related regulations during those years.

The LDA results for the EU show a particular emphasis on clinical evaluation, which aligns with findings from previous research~\cite{mcdermott2024review}, confirming a regulatory focus on clinical aspects in EU documents.

In the USA, terms such as "clinical," "data," and "study" dominate the landscape, reflecting a consistent emphasis on clinical trials, data management, and study protocols. The frequent appearance of "CFR" underscores the reliance on the Code of Federal Regulations, highlighting the importance of codified rules in the regulatory framework. The terms "guidance" and "instructions" appear regularly, indicating ongoing updates and clarifications in regulatory guidelines to adapt to new developments and ensure compliance.

The EU heatmap reveals a strong focus on "clinical," "evaluation," and "data," emphasizing the importance of clinical assessments and data-driven evaluations in the regulatory process. The prominence of terms such as "mdr" (Medical Device Regulation), "udi" (Unique Device Identification), and "European" highlights regulatory initiatives specific to the EU. Additionally, the terms "document," "software," and "security" reflect a focus on documentation requirements, software regulation, and cybersecurity, respectively. The presence of "investigation" and "authority" terms underscores the roles of regulatory investigations and authoritative bodies within the EU's regulatory framework.

In China, the heatmap shows a significant focus on "clinical," "requirements," and "registration," indicating a strong emphasis on clinical requirements and the registration processes for medical devices. The frequent appearance of "data" highlights the importance of data in regulatory processes, while "administration" reflects the role of regulatory administrative bodies. The term "research" underscores the focus on medical research. Additionally, terms such as "adverse," "monitoring," and "technical" indicate a focus on adverse event reporting, post-market surveillance, and technical standards. The presence of "production," "equipment," and "quality" highlights the emphasis on manufacturing standards and quality control within China's regulatory framework.

Comparatively, all three regions consistently emphasize clinical terms, indicating the universal importance of clinical trials and evaluations in medical device regulation. Both the USA and the EU demonstrate a strong focus on "data" and "document," reflecting rigorous data management and documentation requirements. In contrast, China places additional emphasis on "administration" and "requirements," highlighting its regulatory structure and focus on administrative compliance. The EU's focus on terms like "mdr" and "udi" underscores its unique regulatory initiatives, while the USA's consistent reference to "CFR" indicates a reliance on federal regulations. China's emphasis on "administration" and "requirements" further reflects its specific regulatory priorities.

The analysis also highlights the EU's proactive approach to regulating digital health technologies and cybersecurity, as evidenced by the frequent occurrence of terms like "software" and "security." While the USA and China also address these areas, their emphasis appears less frequent compared to the EU.

\begin{figure*}[t]
   \centering
\includegraphics[scale=0.23]{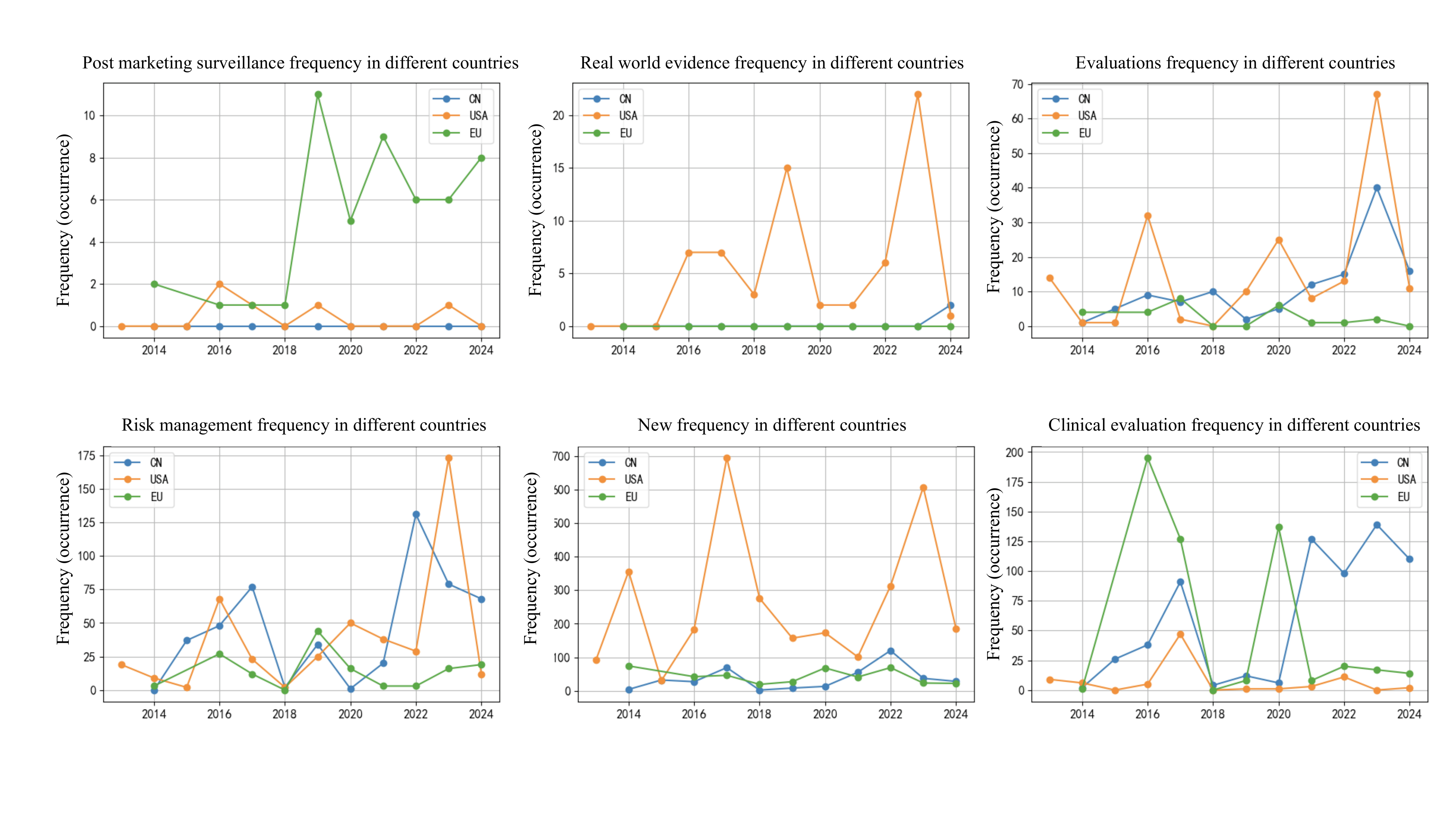}
    \caption{Different keywords frequency over the years}
   \label{fig: keywords}
\end{figure*}

\subsection{Keyword Extraction Across Regulatory Landscapes}

Keyword Extraction is a technique used to automatically identify the most relevant or important words from a text. It enables summarization of the main themes or topics and is commonly applied in information retrieval, content classification, and text summarization tasks.

To enable quick checks of words or themes within a country's regulations and highlight notable trends in regulatory focus and interest, we developed a model called "RegulTermAnalyzer.\cite{yu2024regulterm}" This model performs keyword and key theme frequency analysis of specific terms across different countries. As illustrated in Figure \ref{fig: keywords}, we searched some words, for example, we can see the term "real world evidence" shows significant variation, with the USA exhibiting a prominent peak in 2022, indicating a surge in interest or regulatory emphasis on real world evidence during this period \cite{valla2023use}. This trend is consistent with the USA's increasing integration of real-world evidence into its regulatory framework for medical devices and pharmaceuticals. In contrast, the EU and China show minimal occurrences of this term, suggesting either less emphasis or the use of different terminologies in these regions. According to research, the United States is currently the only country that has explicitly defined Real-World Evidence (RWE) within its formal regulatory framework \cite{valla2023use}. Similarly, the term "evaluations" displays distinct patterns across the regions, with varying frequencies that reflect different regulatory activities and interests over the years. The term "post market surveillance" demonstrates a significant presence in the EU, especially around 2018, indicating a strong focus on monitoring medical devices post-market entry. The USA and CN show lower frequencies, reflecting different regulatory priorities or terminological preferences. The term "new" reveals substantial peaks in the USA, aligning with periods of increased regulatory activity or the introduction of new regulations. Lastly, the frequency of "risk management" shows notable fluctuations, with the USA peaking around 2022, highlighting a heightened focus on regulatory risk management practices during that period . The model developed in this study provides professionals with a valuable tool for examining the frequency and evolution of specific terms within regulatory frameworks of individual countries. It is important to note that the accuracy of the analysis depends on the exact terminology used in the regulations. For instance, querying "real world evidence" will yield accurate results, while using semantically similar terms, such as "true world proof," will not produce correct or relevant data.

\subsection{Vectorization and Comparative Analysis of Corpora}

The vectorization technique is a multi-step process used in Natural Language Processing (NLP) to transform text data into numerical representations and subsequently compare multiple text corpora. This technique involves two main phases, including vectorization and cluster comparison. In vectorization phase, textual data is transformed into vectors, which are numerical representations of words, phrases, or entire documents.

\begin{figure*}[htbp!]
   \centering
\includegraphics[scale=0.07]{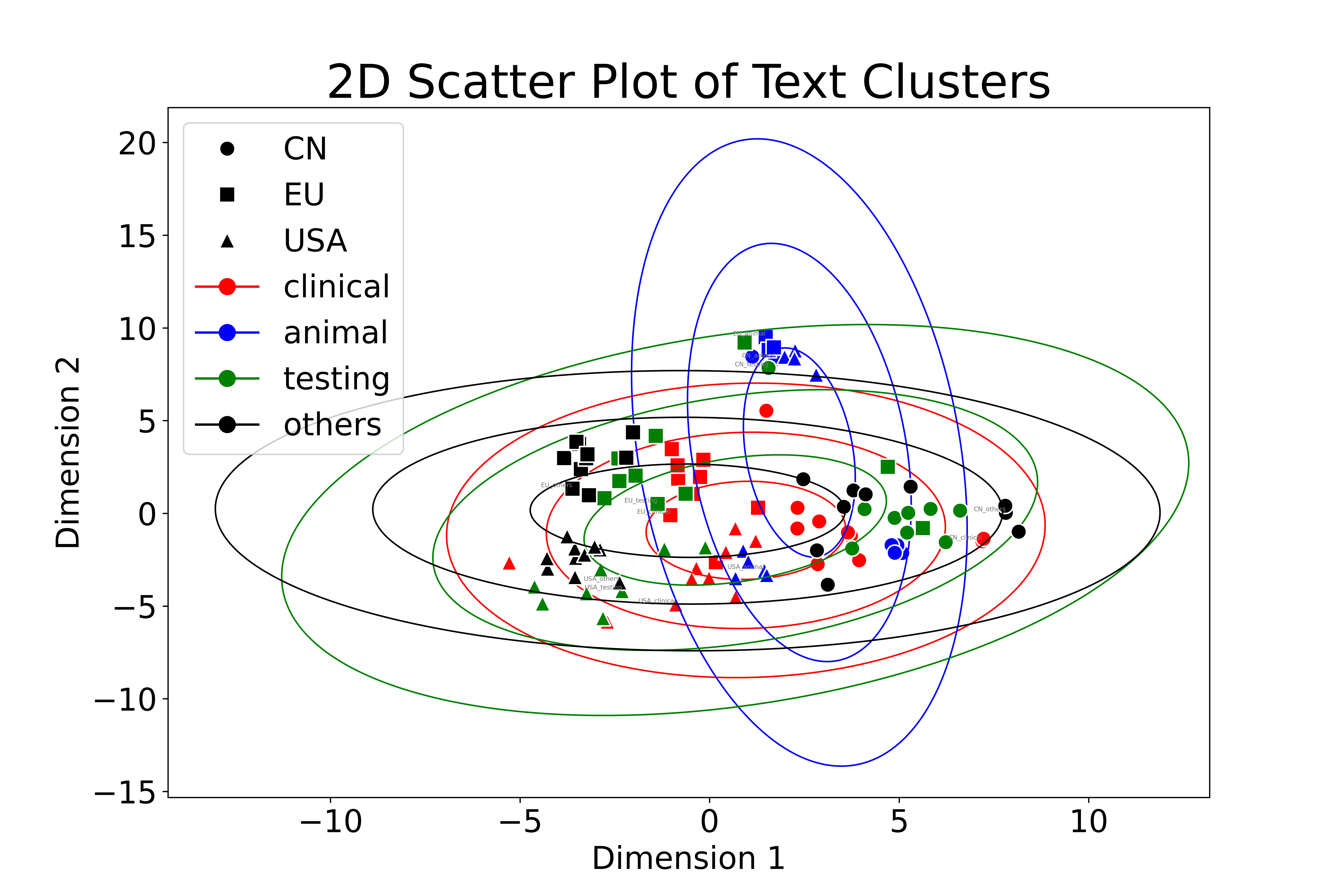}
    \caption{Visualization of 2D Data for NERs Points}
   \label{fig: 2D}
\end{figure*}

\begin{figure*}[htbp!]
   \centering
\includegraphics[scale=0.09]{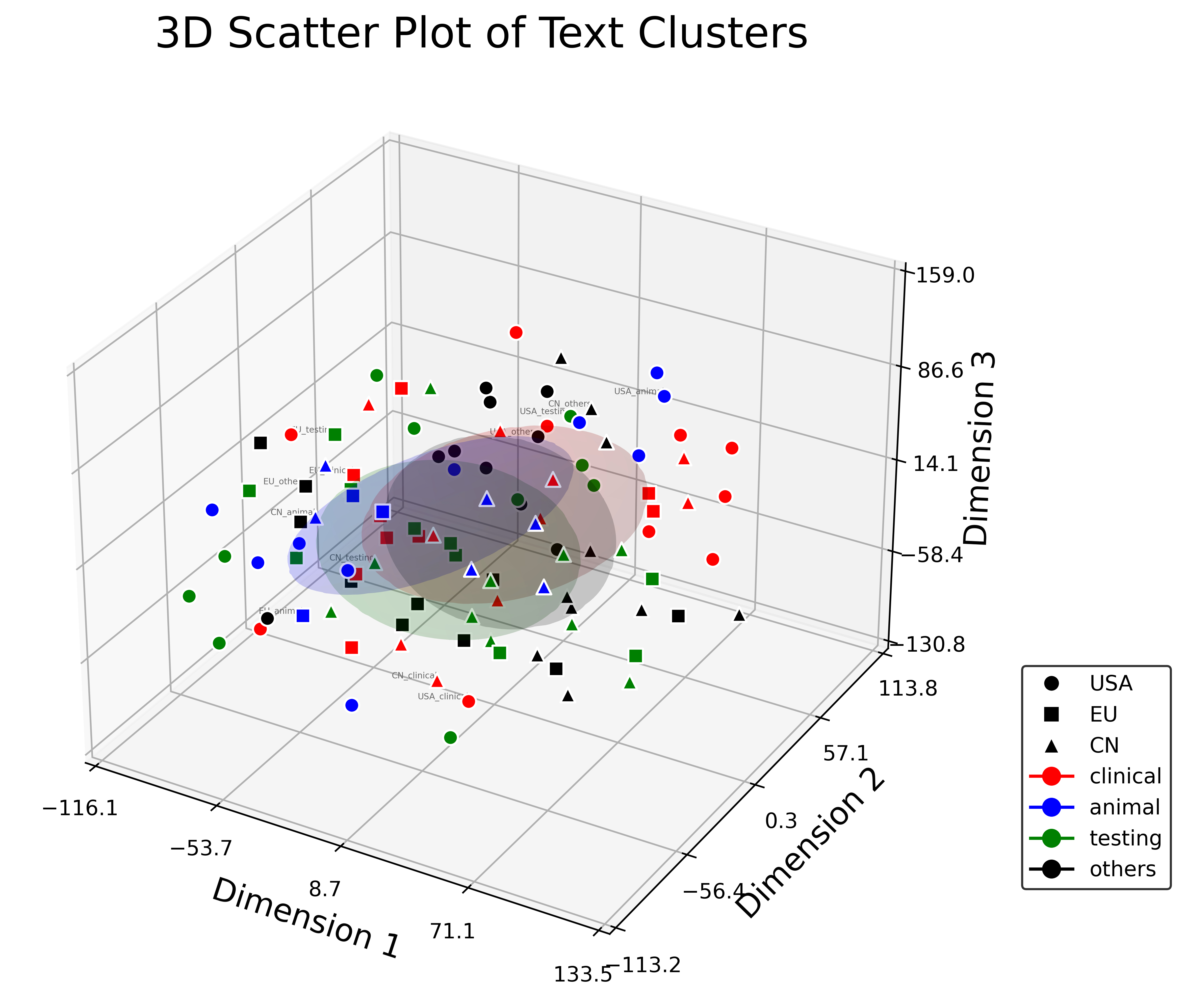}
    \caption{Visualization of 3D Data for NERs Points}
   \label{fig: 3D}
\end{figure*}

\begin{figure*}[htbp!]
   \centering
\includegraphics[scale=0.35]{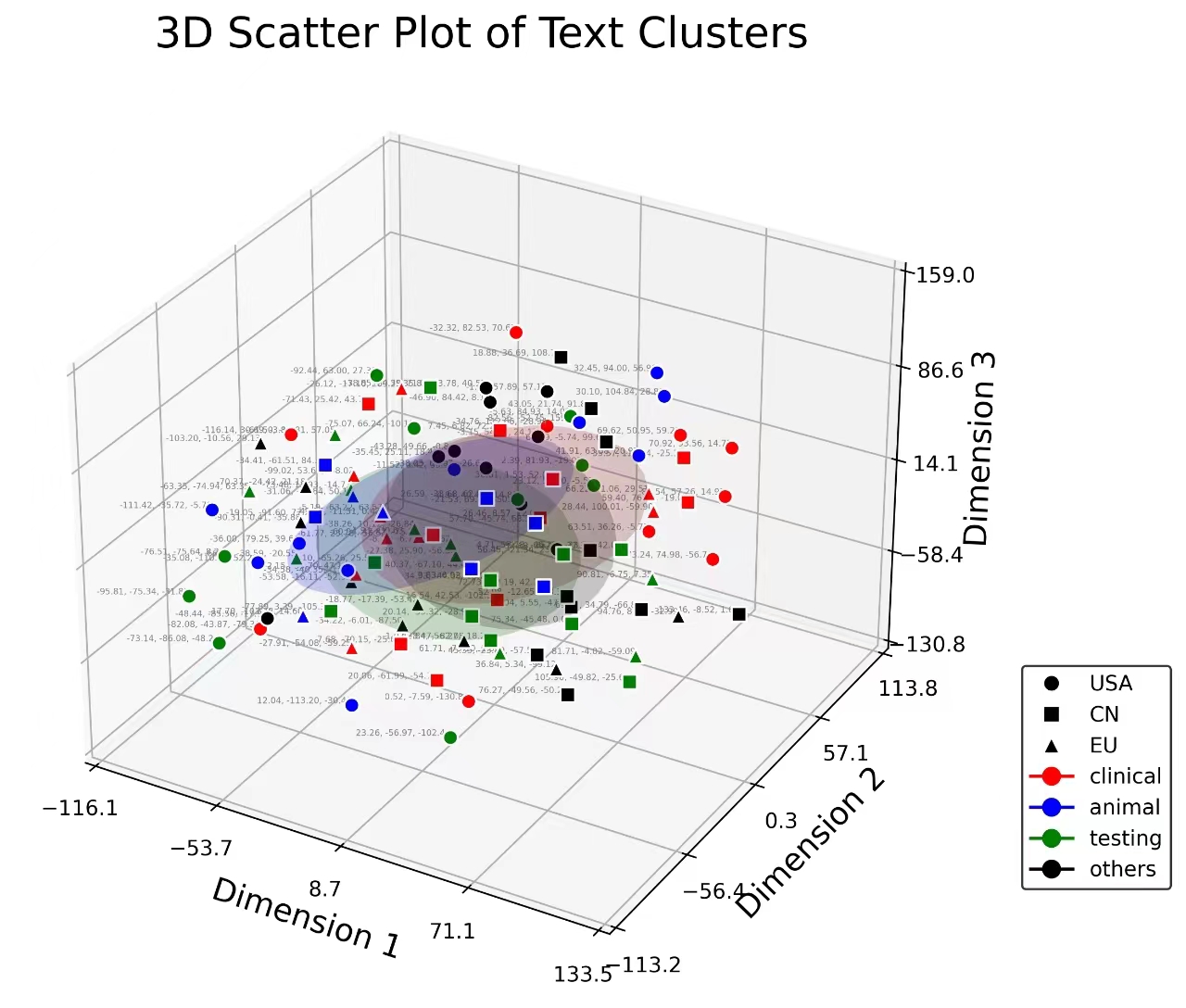}
    \caption{Visualization of 3D Data Points for NERs with Numerical Axes}
   \label{fig: 3Db}
\end{figure*}

As shown in Figure \ref{fig: BERT}, all regulatory documents were pre-processed into individual tokens with multiple labels. The entire corpus was divided into smaller segments, each labeled with three attributes: year, phase (clinical, animal, testing, and other phases), and country. For example, “ClinicalCH2011” represents the Chinese regulatory guidance corpus for the clinical phase of medical devices in 2011. These smaller corpora were then processed for text embedding.

The preprocessed text was then embedded using BERT (Bidirectional Encoder Representations from Transformers), a pre-trained transformer model adept at understanding the context of words within a sentence by considering the surrounding words. Each chunk of text was converted into a high-dimensional vector representation, encapsulating its semantic meaning. Subsequently, the cosine similarity between these text embeddings was computed to assess the semantic similarity of different text chunks. Clustering techniques were then applied to group similar text chunks, providing a visual representation of thematic proximity. After that, we use cosine similarity to measures the cosine angle between two vectors, helping assess how similar the corpora are in terms of content or topic distribution.

We have visualized these corpora using 2D and 3D matrices as shown in Figure \ref{fig: 2D} and Figure \ref{fig: 3Db}. In these visualizations, each point represents a distinct country and a different phase of the process. The results of cosine similarity are shown in Table \ref{tab:3D_cosine_distances}.

\begin{table}[htbp!]
    \centering
    \begin{tabular}{|l|c|c|}
        \hline
        \textbf{Comparison} & \textbf{Mean Cosine Distance} & \textbf{Std Cosine Distance} \\
        \hline
        CN\_animal vs EU\_animal      & 0.718346 & 0.194819 \\
        CN\_animal vs USA\_animal     & 0.326863* & 0.471442 \\
        \hline
        CN\_clinical vs EU\_clinical  & 0.166742 & 0.524871 \\
        CN\_clinical vs USA\_clinical & 0.104228 & 0.503762 \\
        EU\_clinical vs USA\_clinical & -0.089216* & 0.548269 \\
        \hline
        CN\_testing vs EU\_testing    & 0.220312 & 0.483452 \\
        CN\_testing vs USA\_testing   & 0.172389 & 0.568923 \\
        EU\_testing vs USA\_testing   & -0.001876* & 0.530034 \\
        \hline
        CN\_others vs EU\_others      & 0.212934 & 0.544109 \\
        CN\_others vs USA\_others     & 0.052984 & 0.487163 \\
        EU\_others vs USA\_others     & -0.139752* & 0.520475 \\
        \hline
    \end{tabular}
    \caption{3D Mean and Standard Deviation of Cosine Distances for Different Comparisons. Closest pairs are marked with an asterisk (*).}
    \label{tab:3D_cosine_distances}
\end{table}

\noindent
Among all the pairs, we filtered and present the closest pairs as follows in Table \ref{tab:pa}. This is for whole decade corpora together. 

\begin{table}[h!]
    \centering
    \begin{tabular}{l l}
        \toprule
        \textbf{Closest Pair} & \textbf{Mean Cosine Distance} \\
        \midrule
        CN\_animal vs. USA\_animal & 0.326863 \\
        EU\_clinical vs. USA\_clinical & -0.089216 \\
        EU\_testing vs. USA\_testing & -0.001876 \\
        EU\_others vs. USA\_others & -0.139752 \\
        \bottomrule
    \end{tabular}
    \caption{Closest pairs of regulatory text by mean cosine distance.}
    \label{tab:pa}
\end{table}

In Table \ref{tab:pair}, we analyzed regulatory text from the closest pairs identified using mean cosine distance calculated from embedded text vectors. Each pair was condensed into representative chunks, articulating the essential regulatory meanings. We then interpreted the machine-generated similarity by examining the shared regulatory focus across regions. This approach reveals thematic parallels in global regulatory frameworks, enabling a nuanced understanding of common regulatory goals and  interpretation. By presenting the interpretations, the table elucidates why certain points align closely from tons of regulations which is hard to interpret.

\begin{longtable}{|p{2cm}|p{2cm}|p{6cm}|p{4cm}|}
\hline
\textbf{Country} & \textbf{Phase} & \textbf{Details} & \textbf{Shared Focus} \\ 
\hline
\endfirsthead
\hline
\textbf{Country} & \textbf{Phase} & \textbf{Details} & \textbf{Shared Focus} \\ 
\hline
\endhead
\hline \multicolumn{4}{|r|}{\textit{Continued on next page}} \\ \hline
\endfoot
\hline
\endlastfoot

\multicolumn{4}{|c|}{\textbf{Pair 1}} \\ \hline
CN & animal & 
The regulation describes the regulatory requirements for "Preclinical Animal Testing and Clinical Evaluation," stating, “Confirming the composition and structural information of the product is key to determining whether the product can undergo preclinical and clinical research." It specifies that “all preclinical safety studies should be completed before conducting clinical research except for carcinogenicity reproductive and developmental toxicity tests,” which may be exempted based on factors like genotoxicity test results. The passage emphasizes that animal testing should be dosed up to “10 times the highest human dose.” Additionally, it notes that “the requirements for product packaging and packaging integrity should comply with the standard requirements of GBT” and further details guidelines for “preclinical animal testing” and other research materials. The principles for animal testing involve “Replacement, Reduction, and Refinement (3R)” and indicate that non-animal methods should be prioritized when available. & 
Both emphasize preclinical animal testing for medical devices, with allowances for reduced testing if alternative evidence exists. They both stress documentation in animal study reports and ethical testing considerations, with the U.S. using “validated animal models” and China applying the “3R” principles. \\ 
\hline
USA & animal & 
The regulation covers FDA’s regulatory expectations for animal testing. For example, the FDA may require “continued animal testing of implanted devices at 6 months, 1 year, and 2 years after implant.” It specifies that animal study reports must detail “purpose, test method, sample selection, results, discussion of the acceptability of the results, and clinical applicability.” The passage explains that the FDA requires “animal studies to support the initiation of an early feasibility study,” using validated animal models whenever available. It is also noted that when starting pivotal trials, long-term animal studies may be performed concurrently with feasibility studies “to demonstrate complete healing at the implant site.” For certain device evaluations, an animal study may include protocols such as “study design, species, strain, and number of animals used,” and should ensure “anatomic, physiologic, and procedural similarities to humans.” & \\ 
\hline

\multicolumn{4}{|c|}{\textbf{Pair 2}} \\ \hline
EU & Clinical & 
EU regulation outlines the necessary steps for clinical trials, specifying that “groups of subjects rather than individual subjects are allocated” in some trials. It highlights that for these studies, “simplified means for obtaining informed consent will be used.” Details on investigator suitability are provided, including submission of a “current curriculum vitae and other relevant documents,” as well as information on “previous training in the principles of good clinical practice.” Clinical sites need to include descriptions of “the suitability of facilities, equipment, and human resources.” Additional components such as “financial and other arrangements” should also be documented, with details of “compensation paid to subjects and investigator sites.” A substantial modification may be requested for multiple clinical trials by a sponsor, using the EU trial number and “substantial modification code number.” Reporting serious adverse events is mandatory, and unblinding should only occur if “relevant to the safety of the subject.”& 
Both require thorough documentation for clinical trials, including investigator qualifications, ethical considerations, and facilities. They emphasize ongoing safety assessment, with methods suited to each regulatory landscape. \\ 
\hline
USA & Clinical & 
USA regulation discusses FDA’s approach to clinical testing for implanted devices, which may continue “at 6 months, 1 year, and 2 years after implant.” Detailed requirements for individual test reports are specified, including “test method, sample selection, results, and discussion of the acceptability of results.” If pediatric use is intended, additional animal testing might be needed. A long-term animal study is recommended “to demonstrate complete healing at the implant site.” The passage emphasizes that animal testing should occur only when “non-animal testing methods are insufficient.” FDA guidance highlights that the animal study “should involve a validated animal model when available.”& \\ 
\hline

\multicolumn{4}{|c|}{\textbf{Pair 3}} \\ \hline
EU & Testing & 
The notified body must prepare a Clinical Evaluation Assessment Report (CEAR), covering details like device description, intended purpose, and classification. The CEAR assesses the clinical evaluation and equivalence assessments if relevant, clinical investigation plans, and the benefit-risk profile. The notified body verifies that the device meets essential requirements, checking the quality system for procedures in clinical evaluation, risk management, and PMCF (Post-Market Clinical Follow-Up). It ensures clinical data integrity and justifies decisions for compliance with EU directives.& 
Both regulatory processes focus on evaluating and approving medical devices to ensure safety and efficacy. They involve assessments by relevant bodies and collaboration between device and drug manufacturers to streamline development while meeting clinical and safety standards. \\ 
\hline
USA & Testing & 
This guidance assists sponsors in the simultaneous development of antimicrobial drugs and AST devices, aiming to clear AST devices around the same time as new drug approvals. It outlines collaborative efforts between drug sponsors and device makers, covering interactions with both the Center for Drug Evaluation and Research (CDER) and the Center for Devices and Radiological Health (CDRH). The document includes AST device types such as qualitative disc diffusion and other growth-based systems, emphasizing that coordination won’t impact the independent review timelines mandated by MDUFA and PDUFA.& \\ 
\hline

\multicolumn{4}{|c|}{\textbf{Pair 4}} \\ \hline
EU & Others & 
EU regulations, including AIMDD Essential Requirements, demand that any undesirable side effect must represent an acceptable risk when weighed against device performance. The EU emphasizes extensive documentation, such as the Clinical Evaluation Report (CER), which covers device equivalence, risk management, and an ongoing post-market clinical follow-up (PMCF). Furthermore, devices classified for unmet medical needs must comply with Essential Requirements and, in some cases, may be permitted market access with limited clinical evidence if they fulfill significant health benefits. & 
Both frameworks stress rigorous post-market surveillance and documentation to manage risks. They allow expedited pathways for unmet medical needs, aiming to balance public safety and innovation. \\ 
\hline
USA & Others & 
 The USA regulations under 21 CFR outline a rigorous framework for reporting, including 15-day and Five-day Reports for adverse events, malfunction reports, and requirements specific to combination products. Devices must comply with post-market surveillance regulations, including requirements from the Medical Device Reporting (MDR) and the FD\&C Act for tracking safety issues. Additionally, devices classified under specific codes must adhere to performance controls or meet equivalent safety standards through the 510(k) pathway for modifications.& \\ 
\hline
\caption{Similar Points Based on Pairwise Distances and Shared Focus}
\label{tab:pair}
\end{longtable}

\subsection{Cosine Similarity Shifts Over Consecutive Years}

After the IMDRF began releasing guidance documents in 2011, their focus evolved year by year to address critical areas in medical device regulation. In the early years, they concentrated on foundational frameworks in 2013 and continued expanding into clinical evaluations and quality management systems. By 2017, there was a noticeable shift towards advancing clinical evaluations for SaMD, with the IMDRF/SaMD WG/N41 document providing guidance on this emerging technology. Over time, IMDRF documents have increasingly emphasized post-market surveillance and safety, as seen in updates like the Post-Market Adverse Event Reporting Criteria (2023, IMDRF/NCAR WG/N14). We have shown list of regulations in Table \ref{tab:imd}. 

\begin{figure*}[htbp!]
   \centering
\includegraphics[scale=0.5]{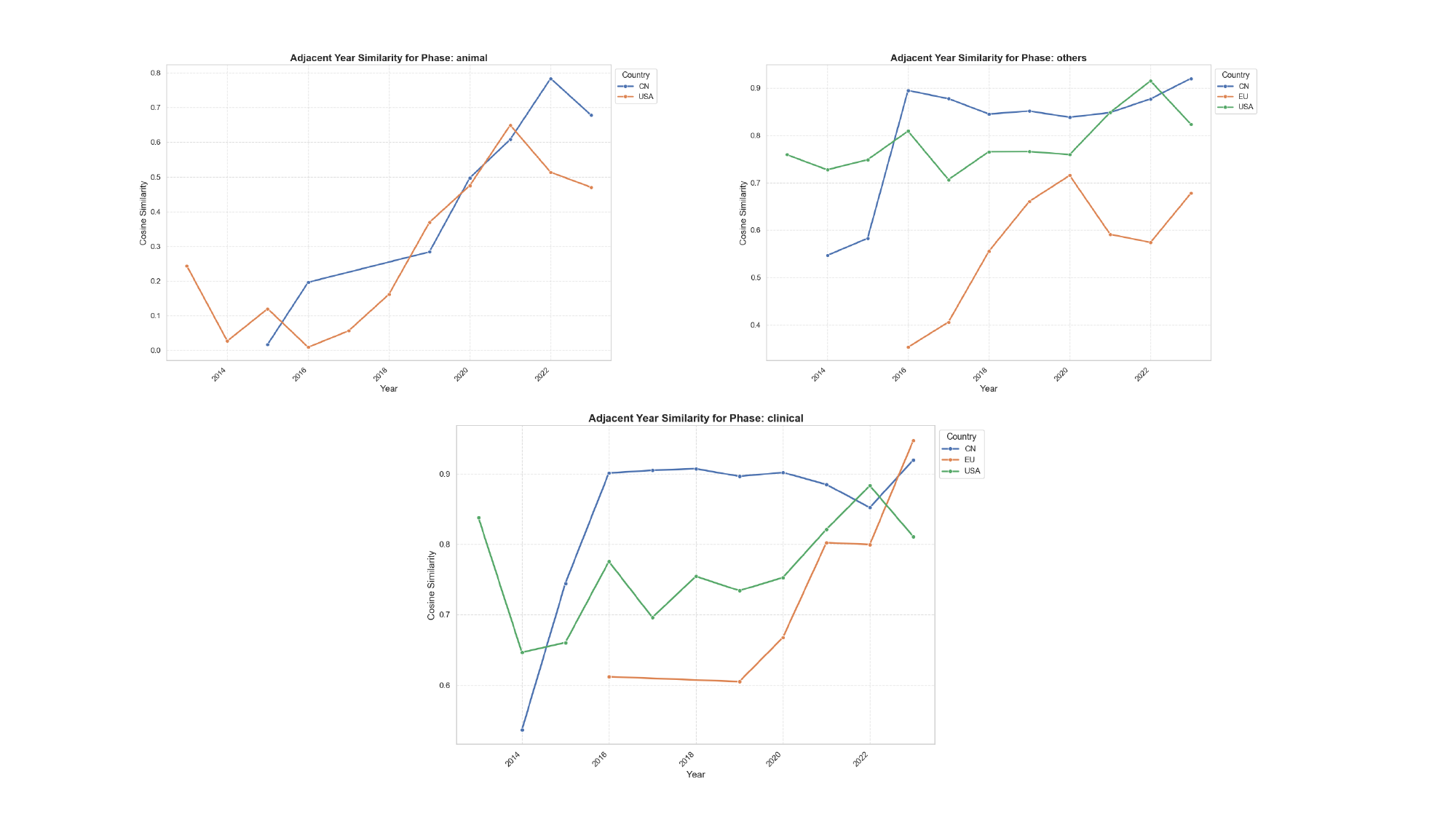}
    \caption{Cosine Similarity Number between Consecutive Years}
   \label{fig: 123}
\end{figure*}

We calculated the cosine similarity number between consecutive years for different phases and countries to assess how consistently the IMDRF's focus shifted or remained stable over time (see Figure \ref{fig: 123}). For example, we visualized the cosine similarity between 2021\_US\_clinical and 2022\_US\_clinical, which helped us identify trends or changes within the same regulatory area year by year. From 2014 onward, there is a noticeable upward trend in similarity values, suggesting that regulations became increasingly aligned from one year to the next. In the animal phase, all countries showed a consistent increase in similarity between consecutive years, indicating that regulations are becoming more harmonized. In the clinical phase, China's similarity values remain consistently high, around 0.9, reflecting a stable and continuous regulatory framework.

\begin{figure*}[htbp!]
   \centering
\includegraphics[scale=0.5]{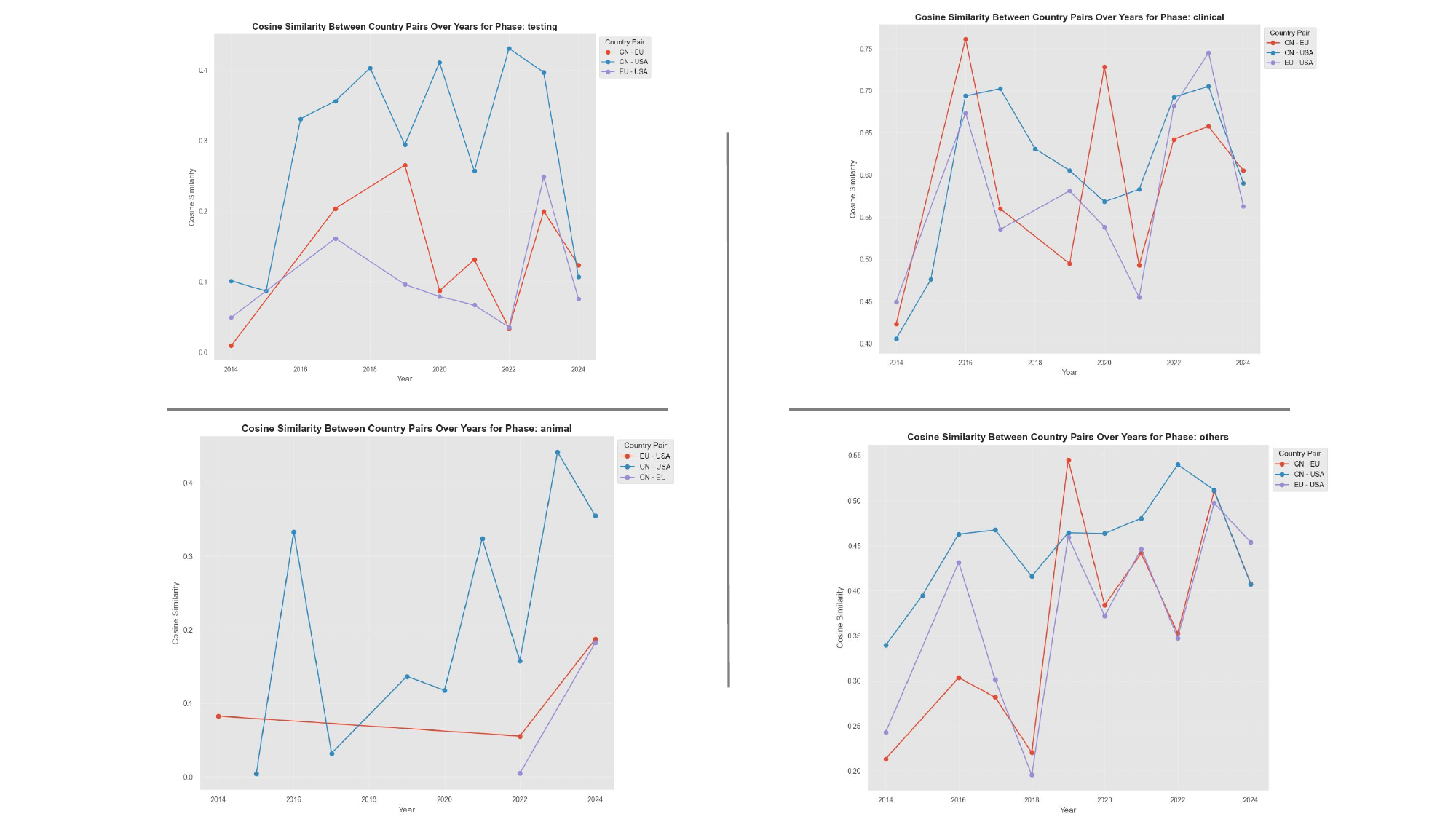}
    \caption{Cosine Similarity Number Between Different Countries}
   \label{fig: 23}
\end{figure*}

A notable exception is the sudden increase in similarity between 2016 and 2017 for China's clinical phase, which aligns with significant regulatory reforms during that time. These reforms included the acceptance of foreign clinical trial data, which reduced the burden on multinational pharmaceutical companies, as well as stricter requirements for clinical trial registration and transparency. Moreover, China’s preparations to join the International Council for Harmonisation of Technical Requirements for Pharmaceuticals for Human Use (ICH) in 2017 likely contributed to this alignment, as the country aimed to standardize its regulations with international norms. These factors explain the sharp rise in similarity between 2016 and 2017. We also compared the corpora between different countries for the same year and phase to explore how regulations vary across regions. This analysis provides insights into how different countries govern the same aspects, such as clinical trials or animal studies, during the same period, highlighting the degree of alignment or divergence between regulatory frameworks (see Figure \ref{fig: 23}). In the figure, we observe that the corpora for testing and animal phases between China and the USA show the highest similarity values in each year.

\begin{table}[htbp]
\centering
\begin{tabular}{|p{5cm}|p{6cm}|p{1cm}|}
\hline
\textbf{Document Name} & \textbf{Description} & \textbf{Year} \\ \hline
Clinical Evaluation (IMDRF/PMD WG/N56) & Framework for assessing clinical data related to the safety and performance of medical devices & 2022 \\ \hline
Post-Market Adverse Event Reporting Criteria (IMDRF/NCAR WG/N14, Edition 4) & Guidelines for reporting adverse events post-market to promote global data sharing & 2023 \\ \hline
Software as a Medical Device: Clinical Evaluation (IMDRF/SaMD WG/N41) & Clinical evidence requirements for software-based medical devices (SaMD) & 2017 \\ \hline
General Principles of Pre-Market Clinical Evaluation (IMDRF/MDCE WG/N47) & Principles for pre-market clinical evaluation of medical devices & 2018 \\ \hline
Optimizing Standards for Regulatory Use (IMDRF/STAND WG/N51) & Best practices for using standards in the regulatory evaluation of medical devices & 2019 \\ \hline
Good Regulatory Review Practices (IMDRF/RPS WG/N49) & Guidance on best practices for medical device regulatory reviews to ensure consistency & 2018 \\ \hline
Definitions for Personalized Medical Devices (IMDRF/PMD WG/N49) & Framework defining key concepts around personalized medical devices & 2020 \\ \hline
Principles for Clinical Evidence (IMDRF/PMD WG/N55) & Outlines requirements for clinical evidence supporting regulatory submissions of personalized medical devices & 2021 \\ \hline
\end{tabular}
\caption{A Series of Publications from IMDRF}
\label{tab:imd}
\end{table}

\subsection{Summarization Clinical Trial Regulations by Region}

Summarizing is a sophisticated technique that leverages the pre-trained language model to create coherent and contextually rich summaries of complex texts.  While sentiment analysis, text summaries, and topic modeling aim to provide a high-level understanding of a document collection, the language summarizing function refines this by generating concise summaries specific to fine-grained regulatory details. 

We employed the SumBERT function to summarize the clinical trial requirements from different regions, as shown in Table \ref{tab:sum}. Given the importance of clinical trials globally, our goal was to see if AI tool, for example, SumBERT can accurately capture key points in each country’s regulations. If works well, this could pave the way for future advancements in digitizing regulatory documents, enabling a more streamlined comparison of international requirements and potentially supporting regulatory harmonization.

\begin{longtable}{|p{2cm}|p{6cm}|p{5cm}|}
\hline
\textbf{Region} & \textbf{Summary of Clinical Trial Requirements} & \textbf{Key Points} \\
\hline
\endfirsthead
\multicolumn{3}{c}{\textit{(continued from previous page)}} \\
\hline
\textbf{Region} & \textbf{Summary of Clinical Trial Requirements} & \textbf{Key Points} \\
\hline
\endhead
\hline
\multicolumn{3}{|r|}{\textit{(continued on next page)}} \\
\hline
\endfoot
\hline
\endlastfoot

European Union (EU) & 
The EU regulations emphasize the autonomy of Member States in managing clinical trial authorizations, ensuring the involvement of ethics committees, and relying on specialized expertise for assessments. However, sponsors may face challenges in providing complete application information across different Member States. & 
\begin{itemize}
\item Member State Responsibility for assessing clinical trial applications.
\item Assessment based on appropriate expertise.
\item Incomplete applications across Member States.
\end{itemize} \\
\hline

China & 
China's regulations focus on the need for localized clinical evaluation data to address specific differences in demographics and clinical practices. If the imported data does not suffice, conducting clinical trials within China is mandatory. Strict compliance with local guidelines for both local and overseas data is required. & 
\begin{itemize}
\item Imported software needs local clinical evaluation data.
\item Mandatory local trials if overseas data is insufficient.
\item Compliance with "Good Clinical Practice" and overseas data guidelines.
\end{itemize} \\
\hline

United States (USA) & 
The USA regulations highlight the importance of comprehensive data collection for certification and disclosure purposes. The FDA emphasizes the potential impact of data from all investigators on study results and requires thorough financial disclosures from clinical investigators. & 
\begin{itemize}
\item Data collection for certification and disclosure.
\item Investigator data can impact overall results.
\item FDA reviews financial disclosure information.
\end{itemize} \\
\hline
\caption{BERTSum Summaries of Country-Specific Clinical Trial Regulations}
    \label{tab:sum}
\end{longtable}

\section{Discussion and Conclusions} \label{Section: conclusion}

The findings from this study reveal significant insights into the regulatory landscapes for medical devices using state of art computational method across China, the USA, and the EU. This section discusses the implications of these findings, the observed trends, and the potential impact on stakeholders in the medical device industry.

\subsection{Implications of LDA Findings}

The application of Latent Dirichlet Allocation (LDA) to the corpus of regulatory documents revealed a nuanced understanding of the key themes and topics that permeate the regulatory landscape for medical devices. This analysis identified distinct areas of focus within each regulatory phase, allowing us to elucidate the specific concerns and priorities that govern the medical device lifecycle.

In the preclinical testing phase, the LDA analysis highlighted topics centered around safety, toxicology, and compliance. These themes underscore the critical nature of thorough evaluations required before medical devices can transition to clinical trials. The emphasis on safety reflects regulatory agencies' commitment to ensuring that devices do not pose undue risks to human subjects. The identification of toxicological assessments indicates a rigorous evaluation of potential harmful effects that materials used in devices may have on biological systems. Compliance discussions further reinforce the necessity for manufacturers to adhere to established regulatory standards and guidelines, thus promoting a culture of safety and accountability in device development.

Moving into the clinical trial phase, the identified topics spotlighted human testing protocols, efficacy, and safety monitoring. These themes reflect the paramount importance of generating robust clinical evidence that demonstrates the safety and effectiveness of medical devices before they can receive regulatory approvals. The focus on human testing protocols indicates the need for rigorous methodologies that protect trial participants while ensuring that the data collected is both reliable and valid. Efficacy discussions highlight the necessity of demonstrating a clear benefit to patients, while safety monitoring signifies the ongoing vigilance required to detect and address any adverse effects that may arise during clinical trials. Collectively, these themes emphasize the intricate balance between advancing innovation in medical devices and safeguarding public health.

The topics emerging from the post-market surveillance phase were predominantly concerned with monitoring device performance, adverse event reporting, and regulatory updates. This phase is crucial as it ensures the continued safety and efficacy of medical devices once they are on the market. The emphasis on monitoring device performance indicates that manufacturers and regulatory bodies must remain vigilant in assessing how devices function in real-world settings. Adverse event reporting is a critical component of this oversight, as it facilitates the identification of potential safety issues that may not have been apparent during clinical trials. Furthermore, discussions surrounding regulatory updates suggest that the medical device landscape is dynamic, necessitating ongoing adjustments to guidelines and practices to respond to new data and emerging technologies. This continuous oversight reflects a commitment to ensuring that medical devices remain safe and effective throughout their lifecycle.

The themes in Table \ref{tab: 5} inferred from the LDA analysis reveal a multifaceted approach to medical device regulation, addressing everything from regulatory frameworks and compliance to clinical and ethical guidelines. Regulatory Framework and Compliance refers to topics like mdr, mdcg, and notified authority highlight the regulatory structures that govern device approval processes. Clinical and Ethical Guidelines refers to the prevalence of terms such as irb, consent, and subjects points to the ethical considerations vital to clinical research. Registration and Diagnostic Equipment refers to keywords such as registration, equipment, and guidelines emphasize the importance of establishing standards for diagnostic tools.

The LDA analysis not only reveals the predominant themes present in the regulatory discourse surrounding medical devices but also underscores the complexity and multidimensionality of the regulatory process. By understanding these topics, stakeholders can better navigate the regulatory landscape, ensuring that the development and approval of medical devices are both efficient and aligned with public health interests.
\subsection{Regional Differences}

The regional comparison highlighted distinct regulatory approaches across China, the USA, and the EU. China's regulatory framework places a strong emphasis on compliance and detailed submission protocols, reflecting a structured and methodical approach to medical device regulation. The USA's focus on clinical trial management and ethical considerations underscores the importance of protecting human subjects and ensuring rigorous clinical evaluation. The EU's balanced focus on safety, efficacy, and data privacy aligns with its broader regulatory environment, particularly the stringent requirements of the GDPR.

These differences have significant implications for medical device manufacturers seeking to enter multiple markets. Understanding the unique regulatory requirements and priorities of each region is essential for successful navigation and compliance. Harmonizing these regulations remains a challenge, but efforts such as the International Medical Device Regulators Forum (IMDRF) are working towards greater alignment and consistency.

The rapid regulatory response to the COVID-19 pandemic demonstrated the ability of regulatory bodies to adapt quickly to global health emergencies. This flexibility is crucial for addressing future public health challenges and ensuring that necessary medical devices can be developed and deployed swiftly \cite{assefa2022attributes}.

For animal studies, the closest pair in the animal studies category is CN\_animal vs USA\_animal, with a mean cosine distance of 0.33. This moderate distance suggests that while there are notable similarities between the animal study regulations in China and the USA, there remains some degree of variability. The relatively higher standard deviation indicates that certain aspects of the regulations may differ significantly, reflecting distinct regulatory approaches or priorities in animal testing protocols. In the clinical studies category, the closest pair is EU\_clinical vs USA\_clinical, with a mean cosine distance of -0.089. Although the negative mean distance indicates a general dissimilarity, this pair is numerically the closest among the comparisons. The high standard deviation suggests substantial variability in clinical trial regulations between the EU and the USA, likely due to different regulatory frameworks and clinical practices. Despite this, the close numerical proximity implies some commonalities, potentially in fundamental principles of clinical trial conduct or evaluation criteria. For testing studies, EU\_testing vs USA\_testing is the closest pair, with a mean cosine distance of -0.002. This near-zero mean distance indicates a high level of similarity between the EU and USA regulations in testing studies. The moderate standard deviation suggests that while the overall regulatory approaches are closely aligned, there may still be specific differences in certain aspects. This alignment could facilitate smoother regulatory processes and mutual recognition agreements between these regions. In the "others" category, the closest pair is EU\_others vs USA\_others, with a mean cosine distance of -0.14. Similar to the clinical studies, the negative mean distance denotes general dissimilarity, but it is the smallest among the comparisons in this category. The standard deviation remains moderate, indicating variability within the regulations. This suggests that while there are differences, there are also shared elements that could form a basis for harmonization efforts.

Results from Figure \ref{fig: 23} and Figure \ref{fig: 123} highlight several turning points, such as China's regulatory reform between 2016 and 2017, which led to a noticeable shift in cosine similarity. This indicates a significant alignment in regulatory documents following the reform. Similarly, the full implementation of the EU Medical Device Regulation (MDR) in 2021 also shows a marked increase in cosine similarity in the clinical phase, reflecting greater consistency in regulatory frameworks following the release of the MDR. These shifts suggest that major regulatory updates lead to more harmonized and structured regulatory content over time.

The analysis highlights significant regional variations in medical device regulations, particularly in clinical and "other" study categories. The closer similarities between the EU and USA across multiple categories point to potential areas for regulatory alignment and cooperation. For China, the moderate similarity with the USA in animal studies suggests an opportunity for harmonization in this specific area. The findings underscore the importance of understanding regional regulatory landscapes to navigate international medical device approval processes effectively. Harmonizing regulations, where feasible, could enhance the efficiency of bringing high-quality medical devices to global markets, ultimately benefiting patients worldwide. Future research could focus on identifying specific regulatory elements that contribute to these similarities and differences, providing a roadmap for achieving greater global regulatory convergence.

\subsection{Implications for Global Regulatory Harmonization}

For stakeholders in the medical device industry, these findings offer valuable insights into the evolving regulatory landscape. Manufacturers must stay informed about regulatory trends and updates to ensure compliance and mitigate risks. The increasing complexity of regulations, necessitates a proactive computational approach to regulatory affairs, including early engagement with regulatory bodies and continuous monitoring of regulatory changes.

Regulatory professionals play a crucial role in bridging the gap between innovation and compliance. By leveraging advanced tools such as NLP for regulatory analysis, they can enhance their ability to manage regulatory requirements and support the successful development and approval of new medical technologies.

The text embeddings, created using BERT, transformed chunks of regulatory text into high-dimensional vector representations that encapsulate the semantic meaning of the content. By computing the cosine similarity between these vectors, we assessed the semantic similarity of different regulatory texts. This process was further enhanced by applying clustering techniques, which grouped similar text chunks and provided a visual representation of thematic proximity. This methodology facilitated a nuanced analysis of the regulatory texts, enabling the identification of patterns and trends within the regulatory landscape.

The findings of this study underscore the complexities and variations in regulatory frameworks across different regions. The closest pairs identified in each phase highlight potential areas for regulatory harmonization. For instance, the moderate similarity between China and the USA in animal studies suggests that these countries could benefit from aligning their regulatory practices to streamline approval processes. Similarly, the high alignment between the EU and USA in testing phases points to the feasibility of mutual recognition agreements that could expedite market access for medical devices.

Despite the insights provided, the study also highlights significant challenges. The high standard deviations across most comparisons indicate substantial variability within regulatory similarities, reflecting the diverse regulatory environments. Future research should focus on deeper analyses of these variations to pinpoint specific regulatory elements that contribute to the observed differences. Additionally, developing automated tools to facilitate real-time updates and comparisons of regulatory frameworks could further enhance the efficiency and efficacy of global medical device regulation.

Overall, this study highlights the dynamic nature of medical device regulations and the critical need for stakeholders to remain adaptable and informed. As the regulatory environment continues to evolve, ongoing research and collaboration between regulators, industry, and academia will be essential for fostering innovation while ensuring the safety and efficacy of medical devices.
\section{Limitation}
This study offers critical insights for researchers, policymakers, and industry professionals, facilitating a more navigable regulatory environment. Through this methodological approach, we aim to contribute to the broader discourse on regulatory convergence, ultimately supporting the IMDRF's mission to create a more efficient and patient-centered global medical device regulatory framework. By leveraging natural language processing (NLP) approaches and other methods within artificial intelligence (AI), our approach can systematically and efficiently analyze large corpora of regulatory documents. This analysis uncovers nuanced insights that are difficult to detect through traditional manual methods. Understanding how Latent Dirichlet Allocation (LDA) topics change over time is particularly useful for manufacturers, as it reveals evolving regulatory priorities and potential future directions. This enables manufacturers to better anticipate regulatory trends, adapt their compliance strategies, and streamline the approval process.

Our study represents the first attempt to use computational method to vectorize regulations, for example, BERT.  However, several limitations should be acknowledged in this analysis. First, the regulations were manually downloaded, which introduces the potential for human error. As a result, some relevant regulations may have been inadvertently omitted due to oversight or accessibility issues. Additionally, the keyword filtering process for the four phases of clinical trials—animal, clinical, testing, and others—may not have been exhaustive. Consequently, there is a risk that some pertinent studies or data points were excluded from our analysis due to limitations in our keyword selection. Plus, there is limitation about NLP on legal or regulatory texts is notoriously challenging due to ambiguity, legal jargon, and region-specific phrasing. 

Moreover, while BERT is a powerful tool for understanding language, it does not fully replicate the nuanced comprehension that a human professional brings to regulatory texts. For instance, in this paper, a human expert noted that if stipulated approval timelines are in place, a sponsor can submit data in multiple rolling review cycles as it becomes available before the formal application submission \cite{ghadanian2024comparison}. However, this process requires more process of coding after the regulation is established, which adds a layer of comprehension work. Current semantic comprehension from regulations, as interpreted by BERT, is therefore less sophisticated than that of a human.

Despite these limitations, this analysis marks a valuable first step in using BERT for regulatory document interpretation. As shown by the similarities in Table \ref{tab:pair}, we demonstrated the model's potential to identify key concepts effectively. Looking ahead, BERT and similar models could play a critical role in supporting researchers, policymakers, and industry professionals by aiding in regulatory harmonization, reducing redundancy, and improving navigation through complex regulatory frameworks. With further refinement and advancements in natural language processing, these tools could offer practical solutions for aligning regulations across jurisdictions and enhancing the efficiency and accuracy of regulatory analysis.

\section{Conflicts of Interest}
The authors affirm that there are no conflicts of interest to declare. 

\bibliographystyle{IEEEtran}
\bibliography{od_prediction}   
\end{document}